\documentclass[12pt]{article}
\usepackage{a4wide,amssymb,cite}
\pdfoutput=1

\usepackage{a4wide,amssymb,graphicx}
\usepackage{epsfig}
\usepackage[usenames,dvipsnames]{color}
\usepackage{slashed}
\usepackage{amssymb,cite,graphicx}
\usepackage{slashed}
\usepackage{amsmath,bm,bbm}
\usepackage{amsfonts}
\usepackage[titletoc,title]{appendix}
\usepackage[small]{caption}
\usepackage[margin=1in]{geometry}
\usepackage[multiple]{footmisc}
\usepackage{mathtools}
\usepackage{slashed}
\usepackage[nottoc]{tocbibind}
\usepackage{xcolor}

\newcommand{\be}{\begin{equation}}
\newcommand{\ee}{\end{equation}}
\newcommand{\bea}{\begin{eqnarray}}
\newcommand{\eea}{\end{eqnarray}}

\def\circa#1{\,\raise.3ex\hbox{$#1$\kern-.75em\lower1ex\hbox{$\sim$}}\,}

\begin{document}

\begin{titlepage}
%
%


%

\begin{centering}
\vspace{1cm}
{\Large {\bf Gravity-Mediated Dark Matter \vspace{0.2cm} \\ at a low reheating temperature}} \\

\vspace{1.5cm}

\begin{centering}
{\bf Hyun Min Lee$^{1,\dagger}$, Myeonghun Park$^{2,\ddagger}$, and Veronica Sanz$^{3,\sharp}$}
\end{centering}
\\
\vspace{.5cm}

{\it $^1$Department of Physics, Chung-Ang University, Seoul 06974, Korea.} \\
{\it $^2$School of Natural Sciences, Seoultech, Seoul 01811, Korea. } \\
{\it  $^3$Instituto de Fisica Corpuscular (IFIC), Universidad de Valencia-CSIC, \vspace{0.02cm} \\ E-46980 Valencia, Spain.}

\vspace{.5cm}


\end{centering}
\vspace{2cm}

\begin{abstract}
\noindent
We present a new study on the Gravity-Mediated Dark Matter (GMDM) scenario, where interactions between dark matter (DM) and the Standard Model are mediated by spin-two particles. Expanding on this established framework, we explore a novel regime characterized by a low reheating temperature, offering an alternative to the conventional thermal relic paradigm. This approach opens new possibilities for understanding the dynamics of the dark sector, encompassing both the dark matter particles (fermion, scalar and vector) and the spin-two mediators.  Our analysis examines the constraints imposed by the relic abundance of DM, collider experiments, and direct detection searches, spanning a wide mass range for the dark sector, from very light to extremely heavy states. This work opens new possibilities for the phenomenology of GMDM.
\end{abstract}

\vspace{2.5cm}

\begin{flushleft} 
$^\dagger$Email: hminlee@cau.ac.kr \\
$^\ddagger$Email: parc.seoultech@gmail.com  \\
$^\star$Email: veronica.sanz@uv.es
\end{flushleft}

\end{titlepage}

\section{Introduction}


The Weakly Interacting Massive Particle (WIMP) paradigm, once considered the leading candidate for dark matter (DM), has faced increasing challenges as experiments have yet to detect any clear signals of WIMPs. This has prompted the need to explore alternative mechanisms for DM production and interaction beyond the traditional thermal freeze-in and freeze-out scenarios. Such mechanisms allow for a broader exploration of the parameter space, accommodating a wider range of dark matter masses, interaction strengths, and production processes.

Recently, a new paradigm has emerged, in which DM is not in thermal equilibrium with the Standard Model (SM) or is heavier than the temperature of the universe at the time of production \cite{lowtemp,Boddy:2024vgt,bernal}. In this scenario, dark matter production relies on freeze-in at low temperatures, where the Boltzmann tail of the SM particle distribution enables the generation of a relic abundance. This approach opens up possibilities for models and parameter spaces that were inaccessible under the assumption of thermal equilibrium.

Gravity-Mediated Dark Matter (GMDM) is particularly intriguing in this context. In GMDM, interactions between DM and the SM are mediated by spin-two particles, which arise naturally in theories with extra dimensions or in certain effective field theory frameworks. A key limitation of the thermal annihilation mechanism in GMDM scenarios is the constraint it imposes on the mediator mass, which requires relatively low masses for the spin-two mediators to ensure efficient annihilation. This limitation motivates the exploration of alternative production mechanisms, which can relax these constraints and allow for heavier spin-two mediators while still accommodating light DM masses.

In this paper, we investigate a new regime for GMDM characterized by a low reheating temperature. This setup provides an alternative to the thermal relic paradigm, resulting in a novel interplay between the properties of the dark sector and the cosmological history. We will demonstrate that this mechanism opens up new possibilities, permitting heavier spin-two mediators without compromising the viability of light DM particles.

This paper is organized as follows. In Section~\ref{sec:GMDM}, we introduce the GMDM framework, while in Sect.~\ref{sec:reheating} we discuss the scenario of low reheating temperature. In Sec.~\ref{sec:DD} we describe the formalism to study the nucleus-Dark Matter scattering with a spin-two mediator and discuss the latest constraints from direct detection experiments. In Sec.\ref{sec:coll} we discuss the different collider signals that this model would produce and which ones are more important for each benchmark scenario. In Section~\ref{sec:results} we present our results, highlighting the viable parameter space and the implications for the dark sector. Finally, in Section~\ref{sec:concl}, we summarize our findings and discuss future directions.

\section{Review of Gravity-Mediated Dark Matter}\label{sec:GMDM}

The Gravity-Mediated Dark Matter (GMDM) \cite{GMDM} framework provides a unique perspective on dark matter interactions with the Standard Model, where spin-two particles mediate between these sectors. This section aims to introduce the theoretical framework of GMDM, focusing on its realization in the context of extra-dimensional theories and its dual interpretation via the AdS/CFT correspondence.

The extra-dimensional picture offers a natural setting for GMDM, with warped geometries such as the Randall-Sundrum model or similar setups providing the necessary structure to localize SM and DM fields. In this framework, the KK excitations of gravitons act as mediators between the SM and a hidden dark sector. These mediators can produce distinct experimental signatures, particularly in colliders, while also influencing the relic abundance and direct detection prospects for DM. 

The dual picture, inspired by the AdS/CFT correspondence, complements the extra-dimensional view by framing the strong coupling dynamics of the dark sector as a dual gravitational description. This perspective not only aids in understanding the phenomenology but also motivates the coupling hierarchies and parameter ranges that emerge in GMDM scenarios.  

In this context, the spin-two mediator is called  the Kaluza-Klein (KK) graviton $G$, although a glueball bound state of new strong interactions would lead to the same phenomenology~\cite{dual}

In the following subsections, we discuss the extra-dimensional realization of GMDM, describing the localization of fields, the effective interactions, and the role of mediator couplings. We then explore the dual interpretation, connecting the gravitational picture to a strongly coupled 4D theory. Together, these complementary views provide a comprehensive foundation for understanding the unique phenomenology of GMDM.

\subsection{The extra-dimensional picture}

In this section, we outline the fundamentals of Gravity-mediated Dark Matter (GMDM) within the context of extra dimensions. Consider a five-dimensional (5D) metric expressed as:
\begin{equation}
ds^2 = w(z)^2 \left(  \eta_{\mu\nu} dx^\mu dx^\nu - dz^2 \right),
\label{metric}
\end{equation}
where \( z \) is the fifth dimension's coordinate, and \( w(z) \) is a smooth function, either decreasing or constant with respect to \( z \). We focus on warped extra dimensions, a choice that will become clearer later on. A well-known example of such warping is found in Anti-de Sitter (AdS) models, like the Randall-Sundrum (RS) scenario~\cite{RS}, where \( w(z) = 1/(kz) \), and \( k \) represents the curvature of the 5D spacetime. The fifth dimension is compactified in the interval \( z \in [0, L] \), with 4D branes positioned at each end. A similar setup can be constructed within a Klebanov-Strassler throat~\cite{KS}. We refer to the brane located at \( z=0 \) as the {\it Matter-brane} and the one at \( L \) as the {\it Dark-brane}.

The effective phenomenology of these models  depend on the localization of bulk  fields\cite{localization}. Fields can be localized on the branes, behaving effectively as four-dimensional fields, or they can live in the full 5D bulk. However, gravity, along with its excitations, propagates throughout the entire 5D spacetime. In our framework, the fields responsible for electroweak symmetry breaking, such as the Higgs boson \( H \) and Dark Matter \( X \), are localized on the Dark-brane. Since Dark Matter's stability and mass are tied to electroweak symmetry breaking, it is natural for these fields to share the same localization. This setup can be realized in models such as the composite Higgs scenario~\cite{CHM}, where \( X \) could belong to the pseudo-Goldstone sector and be protected by residual symmetries~\cite{PomarolDM, chala}. Nevertheless, in the GMDM framework, we explore general dark matter sectors, considering candidates for scalar, vector, and fermion for \( X \).

Meanwhile, gravity and gauge fields propagate through the bulk of the extra dimension, exhibiting 5D dynamics. Massless gauge bosons are delocalized across the bulk with flat profiles, while the gravity mediators, such as the Kaluza-Klein (KK) graviton and the radion, are more localized towards the Dark-brane due to the warping effects.

Instead of proposing a particular origin for \( X \), we focus on its general characteristics: \( X \) is an electroweak-scale singlet under the SM and is stable due to a conserved quantum number driven by the dynamics of the Dark-brane. As an SM singlet, \( X \) interacts with the SM solely via gravitational interactions. The interaction between massless gravitons and any other field is suppressed by \( M_P \), meaning the leading interactions occur through the exchange of other gravitational fields, namely the KK massive gravitons, which we refer to as {\it gravity mediators}. 

The graviton is represented by tensor fluctuations of the metric, with small perturbations introduced into the metric in Eq.~\ref{metric}:
\bea
d s^2= w(z)^2 \left( e^{-2 r} (\eta_{\mu \nu} + G_{\mu\nu} )- (1+ 2 r)^2 d z^2 \right) \ . \label{metric2}
\eea
In this expression, \( G_{\mu\nu} \) is a 5D field that propagate in the extra dimension. We primarily focus on the first Kaluza-Klein (KK) resonance and omit the higher modes. This simplification is valid as long as the higher KK modes are sufficiently separated in mass, and the processes of interest can be represented by interactions involving the lowest KK states of the graviton or the radion. We denote the lowest KK resonances of the graviton  as \( G_{\mu\nu} (x,z) = G_{\mu\nu} (x) f_G(z) \), with \( f_{G}(z) \) representing the wavefunctions along the extra dimension.

The interactions between the SM particles and vector dark matter \( X \) with the KK graviton \( G_{\mu\nu} \) are written in terms of the energy-momentum tensor
\bea
{\cal L}_{\rm KK} = -\frac{1}{\Lambda}G^{\mu\nu}T^{\rm SM}_{\mu\nu} -\frac{1}{\Lambda}G^{\mu\nu} T^{\rm DM}_{\mu\nu}
\eea
More explicitly, the effective interactions of a massive spin-2 field, $G_{\mu\nu}$, to the SM particles and dark matter with arbitrary spin, are given in the following \cite{GMDM},
\bea
{\cal L}_{\rm eff} &=& \frac{c_1}{\Lambda} G^{\mu\nu} \Big( \frac{1}{4} \eta_{\mu\nu} B_{\lambda\rho} B^{\lambda\rho}+B_{\mu\lambda} B^\lambda\,_{\nu} \Big)+\frac{ c_2}{\Lambda} G^{\mu\nu}  \Big( \frac{1}{4} \eta_{\mu\nu} W_{\lambda\rho} W^{\lambda\rho}+W_{\mu\lambda} W^\lambda\,_{\nu}  \Big)  \nonumber \\
&&+ \frac{c_3}{\Lambda} G^{\mu\nu}  \Big( \frac{1}{4} \eta_{\mu\nu} g_{\lambda\rho} g^{\lambda\rho}+g_{\mu\lambda} g^\lambda\,_{\nu} \Big) -\frac{ic_\psi}{2\Lambda}G^{\mu\nu}\left(\bar{\psi}\gamma_{\mu}\overleftrightarrow{D}_{\nu}\psi-\eta_{\mu\nu}\bar{\psi}\gamma_{\rho}\overleftrightarrow{D}^{\rho}\psi\right) \nonumber \\
&&+\frac{c_H}{\Lambda}G^{\mu\nu}\left(2(D_{\mu}H)^{\dagger}D_{\nu}H-\eta_{\mu\nu}\left((D_{\rho}H)^{\dagger}D^{\rho}H-V(H)\right)\right) +\frac{c_{\rm DM}}{\Lambda}\, G^{\mu\nu}  T^{\rm DM}_{\mu\nu}
\eea
where $B_{\mu\nu}, W_{\mu\nu}, g_{\mu\nu}$ are the strength tensors for $U(1)_Y, SU(2)_L, SU(3)_C$ gauge fields, respectively, $\psi$ is the SM fermion, $H$ is the Higgs doublet, and $\Lambda$ is the cutoff scale for spin-2 interactions.
We also note that $c_{i}(i=1,2,3)$, $c_\psi$, and $c_H$ are dimensionless couplings for the massive graviton. Depending on the spin of dark matter, $s=0,\frac{1}{2}, 1$, denoted as $S$, $\chi$ and $X$, we take into account the energy-momentum tensor for dark matter, $T^{\rm DM}_{\mu\nu}$, respectively,
\bea
T^{ S}_{\mu\nu}&=&c_S\bigg[ \partial_\mu S \partial_\nu S-\frac{1}{2}g_{\mu\nu}\partial^\rho S \partial_\rho S+\frac{1}{2}g_{\mu\nu}  m^2_S S^2\bigg],\\
T^{\chi}_{\mu\nu}&=& c_\chi \bigg[\frac{i}{4}{\bar\chi}(\gamma_\mu\partial_\nu+\gamma_\nu\partial_\mu)\chi-\frac{i}{4} (\partial_\mu{\bar\chi}\gamma_\nu+\partial_\nu{\bar\chi}\gamma_\mu)\chi-g_{\mu\nu}(i {\bar\chi}\gamma^\mu\partial_\mu\chi- m_\chi {\bar\chi}\chi) \bigg]
\nonumber \\
&&+\frac{i}{2}g_{\mu\nu}\partial^\rho({\bar\chi}\gamma_\rho\chi)\bigg],  \nonumber \\
T^{X}_{\mu\nu}&=&
c_X\bigg[ \frac{1}{4}g_{\mu\nu} X^{\lambda\rho} X_{\lambda\rho}+X_{\mu\lambda}X^\lambda\,_{\nu}+m^2_X\Big(X_{\mu} X_{\nu}-\frac{1}{2}g_{\mu\nu} X^\lambda  X_{\lambda}\Big)\bigg].
\eea

The couplings of the KK graviton to the SM and dark matter fields depend on their localization within the bulk. In the warped extra-dimension setup, TeV-brane fields have an order one coupling,  e.g. the Dark Matter coupling to gravity-mediators would be, $c_{\textrm{DM}} \simeq {\cal O}(1)$. Field associated with the breaking of electroweak symmetry (the Higgs doublet which contains the Higgs particle and the longitudinal degrees of freedom of massive $W^\pm$ and $Z$) would have a coupling also order 1, $c_H \simeq {\cal O}(1)$.

On the other hand, fields living in the bulk would have a suppressed coupling to gravity-mediators, dictated by the overlap of the wavefunctions in the bulk. For example, gluons and photons (their lowest KK-modes) have a flat profile along the extra-dimension, leading to bulk the volume suppression in their couplings,  $c_{1,2,3} \simeq \frac{1}{\int_{Dark}^{Matter} w(z)} \, d z$. 
Other fields living in the bulk, would have a non-flat profile leading to a smaller overlap with the gravity-mediator, and with coupling given by  $c_\psi = \left(\frac{z_{Matter}}{z_{Dark}}\right)^{\alpha}$, with \( \alpha > 1 \). Moreover, non-trivial boundary conditions of the bulk fields on the branes would lead to possible suppressions of the lowest KK modes, further modifying the spectrum. For example, in Ref.~\cite{cmbV} this feature was used to model the KK couplings to the Standard Model, producing an asymmetry in the CMB spectrum. 

To summarise, the phenomenology of fields embedded into an extra-dimension is very rich, and depends strongly on the geometry of the extra-dimension, localization and boundary conditions. Nonetheless, there are some generic features one can extract, like the flatness of gauge bosons or the quasi-localization of massive states in the bulk. In Sec.\ref{sec:dual} we will discuss how these features have a holographic description in terms of a strongly coupled new sector, but before moving into the dual picture, we provide two useful benchmarks which will help us explore a range of possibilities for GMDM.

\subsubsection{Coupling Benchmarks}

To explore the phenomenology of the GMDM scenario, we consider two distinct benchmarks for the couplings of the spin-two mediator to Standard Model and DM fields: the universal case and the bulk case. These benchmarks illustrate how different assumptions about the localization of fields and the structure of the extra dimension impact the interactions between the graviton, DM, and SM particles.

{\bf $\bullet$ The universal case: } In the case that all SM and DM fields are localized on the 4D TeV-brane, all couplings would be equal, 
\begin{equation}
   \textrm{\bf Universal case: }c_{DM}= c_{1,2,3}=c_H=c_{\psi} \, 
\end{equation}
where $\psi$ denotes any SM fermion and the indices $i=$3, 2 and 1 refer to the coupling to the gauge forces $SU(3)_c \times SU(2)_L \times U(1)_Y$, respectively. 

\begin{figure}[t!]
    \centering
\includegraphics[width=0.95\linewidth]{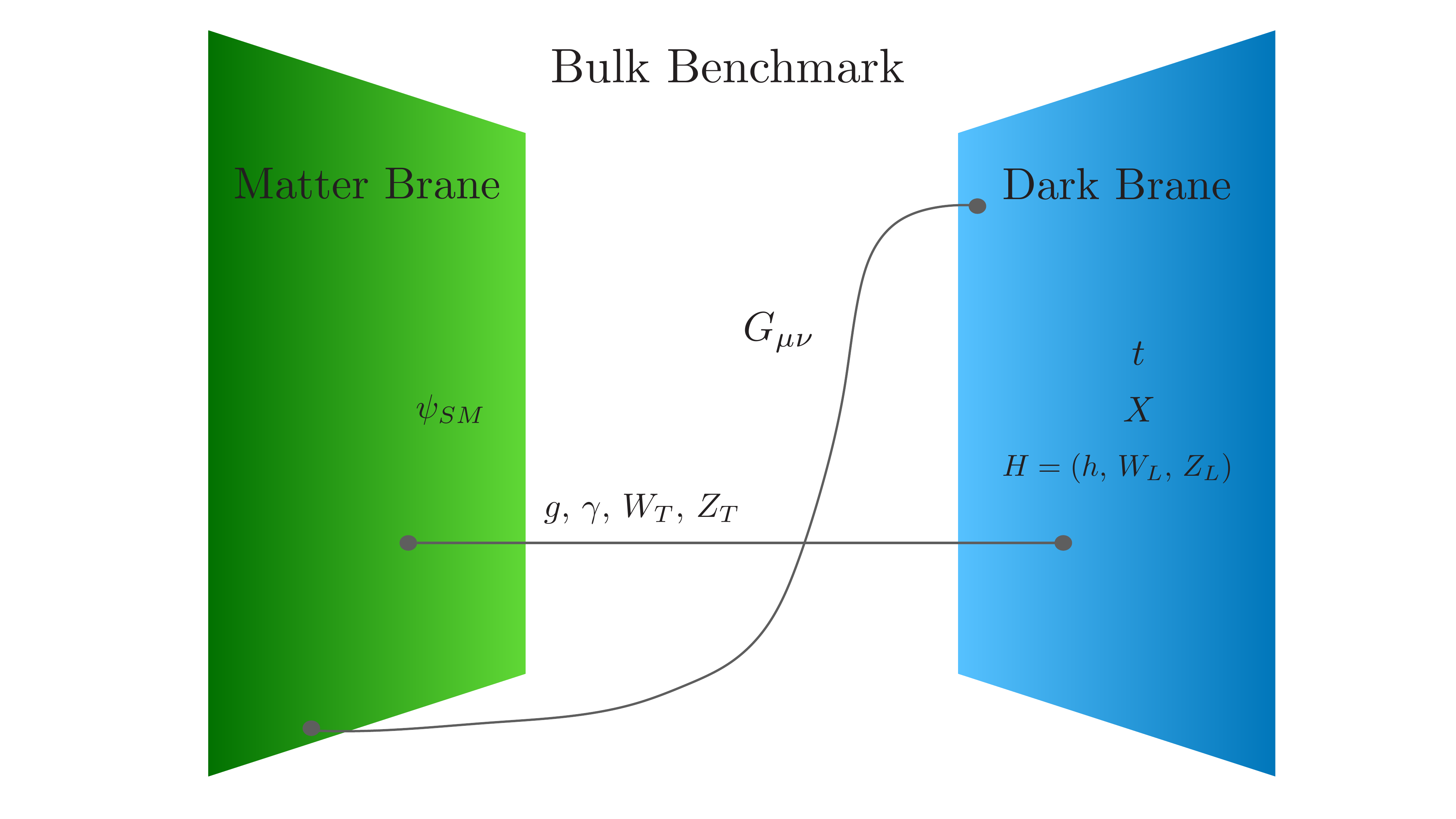}
    \caption{Schematic depiction of the bulk benchmark scenario.}
    \label{fig:bulkbench}
\end{figure}

{\bf $\bullet$ The bulk case: } In this benchmark we consider models where the origin of mass may be related to the extra-dimensional dynamics, see e.g. Ref.~\cite{RSbulk} for more details. In this case, one can use the bulk as a way to localize fields and create hierarchies among the couplings of the graviton to SM fields. A particularly interesting benchmark corresponds to locating near or on the Dark Brane the DM field and the field responsible of electroweak symmetry breaking. Massless gauge fields, if placed in the bulk, will have a flat profile, whereas fermionic bulk fields can be localized using a bulk mass parameter. See Fig.~\ref{fig:bulkbench} for a schematic depiction of this scenario.

For concreteness, let us focus on warped AdS models. The KK-graviton $G$ is localized nead the Dark brane at $z=L$ and all the other fields which are on that brane or localiized near-by will have order one couplings to the graviton, hence
\begin{equation}
    c_{H} \simeq c_{DM} \simeq {\cal O}(1) \,
\end{equation}
where one needs to remember that, after electroweak symmetry breaking, part of the $W$ and $Z$ will reside in the Higgs doublet $H$ and will hence exhibit order one couplings to the graviton.

The  gauge fields, on the other hand, will have a flat profile in the bulk and its coupling will then be suppressed by the partial overlap of this flat profile with a graviton $G$ localised near the Dark Brane. In AdS metrics, the value of \( c_{1,2,3} \) can be expressed as 
\begin{equation}
    c = 2 \frac{1-J_0(x_G)}{\log\left(\frac{M_{Pl}}{TeV}\right) \,  x_G^2 \, |J_2(x_G)|},
\end{equation}
where \( x_G = 3.83 \) is the first zero of the Bessel function \( J_1 \), in the absence of localized kinetic terms. This expression also shows the suppression by $1/\log(M_P/TeV) \sim {\cal O}(0.03)$. 

On the other hand, fermion coupling  $c_\psi$ would be given by
\begin{equation}
    c_\psi = \sqrt{\frac{3}{2}} \,  \frac{1+2 \nu_\psi}{1-e^{-k L (1+2 \nu_\psi)}} \, \int_0^1 d y y^{2+2 \nu_\psi} \frac{J_2(3.83 y)}{J_2(3.83)},
\end{equation}
where \( \nu_\psi = M_\psi / k \) is a dimensionless mass parameter, with \( k \) being the AdS curvature and \( M_\psi \) the bulk fermion mass~\cite{RSbulk}. 

When \( \nu_\psi = 1/2 \), the fermion zero mode is flat in the extra dimension, referred to as the {\it conformal value}. For \( \nu_\psi > 1/2 \), the fermion zero-mode localizes near the Dark-brane, while for \( \nu_\psi < 1/2 \), it localizes closer to the Matter-brane. This bulk mass parameter has a dual interpretation in the context of partial compositeness~\cite{fdual1}. In practice, all light fermions will have a negligible coupling to the graviton. Hence we will just take into account the coupling to tops.

In summary, the bulk case is characterized by:
\begin{equation}
    \textrm{\bf Bulk case: } c_{DM} \simeq c_{H} \simeq c_{t} \simeq (kL) c_{g,\gamma} \gg c_{f},
\end{equation}
where for the typical Randall-Sundrum model one expects $kL\simeq {\cal O}(50)$. Also note that here  $c_f$ refers to the couplings to light fermions.

\subsection{The dual picture}\label{sec:dual}

The AdS/CFT correspondence has been extensively studied~\cite{AdSCFTgen}, highlighting a duality between strongly coupled theories in $D$ dimensions and a gravitational theory in $D+1$ dimensions. This duality extends beyond supersymmetric or strictly conformal theories~\cite{LisaPoratti}. Broadly, this holographic relationship connects strongly coupled systems (the target theory) with a higher-dimensional theory, often considered an {\it analogue computer}~\cite{analogue}, offering improved calculability. 

This section discusses the holographic dual description of the model introduced earlier. In this dual picture, the bulk of the extra dimension encodes the RG flow of the 4D Lagrangian, with the Matter-brane and Dark-brane serving as the UV and IR boundary conditions, respectively. Moving from the Matter-brane to the Dark-brane corresponds to integrating out degrees of freedom. At a given position $z_*$, the {\it local} cutoff is related to the UV cutoff as~\cite{lisa-matt}
\bea
\Lambda (z_*) = \omega(z_*) \Lambda_{UV} \ .
\eea

The Dark-brane signals the confinement of a sector of the theory, leading to the appearance of {\it composite} states (Kaluza-Klein modes) at low energies and their localization near the Dark-brane. Conversely, fields localized on or near the Matter-brane do not strongly participate in the dynamics near the Dark-brane and are considered {\it elementary}. Thus, the localization of fields reflects their degree of compositeness.

Gauge fields that are de-localized (flat) in the extra dimension represent global symmetries of the composite sector, weakly gauged by the UV dynamics~\cite{csaba-fat}. These fields are mixtures of composite and elementary components, analogous to $\rho-\gamma$ mixing in QCD~\cite{ami-qcd,alex-qcd,QCD-hol}.

From the dual perspective, gravity mediators have a specific interpretation. Their presence reflects the conformal symmetry of the composite sector, which is spontaneously broken by strong dynamics. 

The massive KK graviton is linked to a broken diffeomorphism invariance of the composite sector in 4D. Without breaking, the conserved massless spin-two field $\theta^{\mu\nu}$ satisfies $\partial_{\mu} \theta^{\mu\nu}=0$. Breaking diffeomorphism invariance leads to $\partial_{\mu} \theta^{\mu\nu} = a^{\nu}$, where $a^{\nu}$ is a massive vector field that the spin-two field {\it eats}, forming a massive spin-two state $\tilde{G}$~\cite{massives2}. The massive spin-two field couples to particles as:
\bea
- \frac{c_i}{\Lambda_{\tilde{G}}} \, \tilde{G}_{\mu\nu} \, T^{\mu\nu}_i \ ,
\eea
where $\Lambda_{\tilde{G}}$, represents the scale of spontaneous breaking of the symmetry.

In summary, the holographic dual of our warped extra-dimensional model illustrates partial compositeness. Gravity mediator reflect the breaking of conformal symmetry at scale $\Lambda$. Dark-brane states are fully composite, Matter-brane states are elementary, and bulk gauge fields exhibit partial compositeness, as their gauge bosons arise from weakly gauging global symmetries of the composite sector.

\section{Dark matter freeze-in at a low temperature}
\label{sec:reheating}

In this section, we discuss the production of dark matter (DM) through freeze-in at a low reheating temperature, a mechanism that relies on DM not being in thermal equilibrium with the Standard Model (SM). This scenario opens new avenues for dark matter models, particularly those where interactions are mediated by heavy particles, such as spin-two gravitons in the GMDM framework. 

At low temperatures, the production of dark matter occurs through the Boltzmann tail of the SM particle distribution, allowing otherwise forbidden processes to generate DM particles. This approach is particularly relevant when the reheating temperature is much smaller than the DM mass, as thermal equilibrium cannot be established. 

We first derive the relevant Boltzmann equations governing DM production, focusing on the contributions from annihilation processes involving SM particles. Then, we calculate the resulting relic density for various DM candidates, including fermions, vectors, and scalars, under the freeze-in framework. Finally, we highlight how the freeze-in mechanism depends on the reheating temperature, interaction cross-sections, and mediator properties.

\subsection{Boltzmann equations}

For Dirac fermion dark matter, assuming that there is no asymmetry in the dark matter abundance, we set $n_{\chi}=n_{\bar\chi}$. Then, the number densities, $n_\chi$ and $n_{\bar\chi}$, are determined by the same Boltzmann equation,
\bea
{\dot n}_\chi + 3H n_\chi =-\langle \sigma v\rangle_{{\bar\chi}\chi\to {\bar f}f} n^2_\chi + \langle\sigma v\rangle_{{\bar f}f\to {\bar\chi}\chi} (n_{f,{\rm eq}})^2.
\eea
Then, the corresponding Boltzmann equation for the total number density of fermion dark matter, $n_{\rm DM}=n_\chi+n_{\bar\chi}=2n_\chi$, is given by
\bea
{\dot n}_{\rm DM} + 3H n_{\rm DM} =-\frac{1}{2} \langle \sigma v\rangle_{{\bar\chi}\chi\to {\bar f}f} n^2_{\rm DM} + 2\langle\sigma v\rangle_{{\bar f}f\to {\bar\chi}\chi} (n_{f,{\rm eq}})^2.
\eea
Here, we considered only ${\bar f}f\to {\bar\chi}\chi$ and its inverse process, but we can include other creation and annihilation channels for dark matter, except $GG\to {\bar\chi}\chi$ and its inverse process, which are negligible for $m_G\gtrsim m_\chi$. 
Moreover, we can recast the above Boltzmann equation to the equation for the dark matter abundance, $Y=\frac{n_{\rm DM}}{s}$, with $x=m_\chi/T$, as
\bea
\frac{dY}{dx} = \lambda x^{-2} \bigg(-\frac{1}{2} \langle \sigma v\rangle_{{\bar\chi}\chi\to {\bar f}f} Y^2 + 2\langle\sigma v\rangle_{{\bar f}f\to {\bar\chi}\chi} (Y_{f,{\rm eq}})^2 \bigg), \label{Boltzmann}
\eea
with $\lambda= s(m_\chi)/H(m_\chi)$ and $s(m_\chi)=\frac{2\pi^2}{45} g_{*s} m^3_\chi$ and $H(m_\chi)=\sqrt{\frac{\pi^2 g_*}{90}}\, \frac{m^2_\chi}{M_P}$.

At a low temperature, $T\ll m_\chi$, the dark matter production, ${\bar f}f\to {\bar\chi}\chi$, is forbidden, so we need to rely on the Boltzmann tail of the number density of the SM particles with $E_f, E_{\bar f}\gtrsim m_\chi$. 
In order to obtain the dark matter production rate in the Boltzmann equation, we use the following detailed balance condition,
\bea
2\langle\sigma v\rangle_{{\bar f}f\to {\bar\chi}\chi} (n_{f,{\rm eq}})^2=\frac{1}{2} \langle \sigma v\rangle_{{\bar\chi}\chi\to {\bar f}f} (n_{\rm DM, eq})^2,
\eea
which gives rise to
\bea
\langle\sigma v\rangle_{{\bar f}f\to {\bar\chi}\chi}= \frac{(n_{\rm DM, eq})^2}{4n^2_{f,{\rm eq}}}\, \langle \sigma v\rangle_{{\bar\chi}\chi\to {\bar f}f}. \label{balance}
\eea
Thus, it is sufficient to know the annihilation cross section, $\langle \sigma v\rangle_{{\bar\chi}\chi\to {\bar f}f}$, with the same total energy, $E_{\chi}+E_{\bar\chi}=E_f+E_{\bar f}$, as in the production process for dark matter, ${\bar f}f\to {\bar\chi}\chi$.

As a result, using eq.~(\ref{balance}), we can rewrite eq.~(\ref{Boltzmann}) as
\bea
\frac{dY}{dx} = \frac{1}{2}\lambda x^{-2}\langle \sigma v\rangle_{{\bar\chi}\chi\to {\bar f}f} (Y^2_{\rm eq} - Y^2 ), \label{Boltzmann2}
\eea
with
\bea
Y_{\rm eq}= \frac{45 g_{\rm DM}}{ 4\pi^4 g_{*s}}\, x^2 K_2(x).
\eea
Here, $K_2(x)$ is the modified Bessel function of the second kind, and $g_{\rm DM}$ is the number of degrees of freedom for fermion dark matter, which is $g_{\rm DM}=4$.

\subsection{Dark matter relic density}

Assuming that fermion dark matter is not in thermal equilibrium with the SM bath, we can ignore the $Y^2$ term on the right-hand side of eq.~(\ref{Boltzmann2}).
Then, parametrizing $\langle \sigma v\rangle_{{\bar\chi}\chi\to {\bar f}f}=\sigma_0 x^{-n}$ and integrating the Boltmann equation in eq.~(\ref{Boltzmann2}) between $x_R=m_\chi/T_{\rm RH}$  and $x=\infty$, with $T_{\rm RH}$ being the reheating temperature, we obtain the dark matter abundance at $x=\infty$ as
\bea
Y_\infty = \frac{1}{2} \sigma_0 \lambda\int^\infty_{x_R} dx\, x^{-n-2}\, Y^2_{\rm eq}.
\eea
Then, approximating $K_2(x)\simeq \sqrt{\frac{\pi}{2x}}\,e^{-x}$ for $x=m_\chi/T\gg 1$, we obtain 
\bea
Y_\infty \simeq \frac{\pi}{8} \bigg(\frac{45g_{\rm DM}}{4\pi^4 g_{*s}}\bigg)^2\bigg(\sqrt{\frac{90}{\pi^2}} \frac{g_{*s}}{g^{1/2}_*}\, m_\chi M_P\sigma_0\bigg)\cdot e^{-2m_\chi/T_{\rm RH}}\, \bigg(\frac{m_\chi}{T_{\rm RH}}\bigg)^{1-n}.
\eea
Here, we note that
\bea
\int^\infty_{x_R} dx\,  x^{-n+1}\, e^{-2x} = 2^{-2+n} \Gamma(2-n,2x_R)\simeq \frac{1}{2}\, e^{-2x_R}\, x^{1-n}_R, \quad x_R\gg 1.
\eea

Finally, the relic density for dark matter is given by
\bea
\Omega_{\rm DM} h^2 = 0.2745\bigg( \frac{Y_\infty}{10^{-11}}\bigg) \bigg(\frac{m_\chi}{100\,{\rm GeV}}\bigg).
\eea

\subsection{Dark matter with other spins}

When dark matter is a real vector or scalar field, we can consider the DM annihilation processes, $XX\to f{\bar f}$ and $SS\to f{\bar f}$. 
In this case, the Boltzmann equations for the number density of dark matter, with $B=X,S$, are
\bea
{\dot n}_B +3 H n_B = -2\langle\sigma v\rangle_{BB\to f{\bar f}} n^2_B + 2\langle\sigma v\rangle_{f{\bar f}\to BB} (n_{f,{\rm eq}})^2.
\eea
We note that we can include other creation and annihilation channels for dark matter, except $GG\to {\bar\chi}\chi$ and its inverse process, which are negligible for $m_G\gtrsim m_B$. 
Then, for $Y_B=\frac{n_B}{s}$ and $x=m_B/T$, the above equation can be recasted into
\bea
\frac{dY_B}{dx}=\lambda x^{-2} \bigg(  -2\langle\sigma v\rangle_{BB\to f{\bar f}} Y^2_B + 2\langle\sigma v\rangle_{f{\bar f}\to BB} (Y_{f,{\rm eq}})^2\bigg). \label{YB}
\eea
In this case, similarly to the case of fermion dark matter, the detailed balance condition  is
\bea
\langle\sigma v\rangle_{BB\to f{\bar f}} (n_{B,{\rm eq}})^2 = \langle\sigma v\rangle_{f{\bar f}\to BB} (n_{f,{\rm eq}})^2,
\eea
which leads to
\bea
 \langle\sigma v\rangle_{f{\bar f}\to BB} =\frac{(n_{B,{\rm eq}})^2 }{ (n_{f,{\rm eq}})^2}\, \langle\sigma v\rangle_{BB\to f{\bar f}}. 
\eea
Thus, eq.~(\ref{YB}) can be rewritten as
\bea
\frac{dY_B}{dx} = 2\lambda x^{-2 }\langle \sigma v\rangle_{{\bar\chi}\chi\to {\bar f}f} ((Y_{B,\rm eq})^2 - Y^2_B ), \label{YB2}
\eea
with
\bea
Y_{B,{\rm eq}}= \frac{45 g_B}{ 4\pi^4 g_{*s}}\, x^2 K_2(x)
\eea
where $g_B=3, 1$  for real vector and scalar dark matter, respectively.

Therefore, ignoring the $Y^2_B$ term  in eq.~(\ref{YB2}) and taking $ \langle\sigma v\rangle_{BB\to f{\bar f}}=\sigma_B\, x^{-n}$, we determine he abundance for vector or scalar dark matter by
\bea
Y_{B,\infty}=2\sigma_B \lambda\int^\infty_{x_R} dx\, x^{-n-2}\, (Y_{B,{\rm eq}})^2,
\eea
which is approximated for $x_R\gg 1$ as
\bea
Y_{B,\infty} \simeq \frac{\pi}{2} \bigg(\frac{45g_B}{4\pi^4 g_{*s}}\bigg)^2\bigg(\sqrt{\frac{90}{\pi^2}} \frac{g_{*s}}{g^{1/2}_*}\, m_B M_P\sigma_B\bigg)\cdot e^{-2m_B/T_{\rm RH}}\, \bigg(\frac{m_B}{T_{\rm RH}}\bigg)^{1-n}.
\eea

\section{Direct Detection}
\label{sec:DD}


Direct detection experiments aim to observe dark matter (DM) through its elastic scattering off nuclei, mediated by spin-two gravitons in the Gravity-Mediated Dark Matter (GMDM) framework. Using the results from Ref.~\cite{GMDM-ddnew}, the total cross-section for spin-independent elastic scattering between dark matter and a nucleus is given by
\bea
\sigma_{{\rm DM}-A}^{SI} = \frac{\mu_A^2}{\pi}\,\Big(Z f^{\rm DM}_p+(A-Z) f^{\rm DM}_n\Big)^2,
\eea
where $\mu_A = m_\chi m_A/(m_\chi + m_A)$ is the reduced mass of the DM-nucleus system, $m_A$ is the target nucleus mass, and $Z$ and $A$ are the number of protons and the atomic number, respectively.

The nucleon form factors, $f^{\rm DM}_p$ and $f^{\rm DM}_n$, capture the effective couplings of DM to protons and neutrons, respectively. These are given by the same expressions for all spins of dark matter:
\bea
f^{\rm DM}_p &=& \frac{c_{\rm DM} m_N m_{\rm DM}}{4m_G^2 \Lambda^2}\bigg(\sum_{q=u,d,s,c,b} 3 c_q \big(q(2)+\bar{q}(2)\big) + 3 c_g G(2) \nonumber \\
&&+ \sum_{q=u,d,s} \frac{1}{3} c_q \big(f^p_{Tq} - \frac{2}{27} f_{TG}\big) + \frac{11}{9} c_g f_{TG}\bigg) \nonumber \\
&\equiv& \frac{c^p_{\rm eff} c_{\rm DM} m_N m_{\rm DM}}{4m_G^2 \Lambda^2}, \label{fp} \\
f^{\rm DM}_n &=& \frac{c_{\rm DM} m_N m_{\rm DM}}{4m_G^2 \Lambda^2}\bigg(\sum_{q=u,d,s,c,b} 3 c_q \big(q(2)+\bar{q}(2)\big) + 3 c_g G(2) \nonumber \\
&&+ \sum_{q=u,d,s} \frac{1}{3} c_q \big(f^n_{Tq} - \frac{2}{27} f_{TG}\big) + \frac{11}{9} c_g f_{TG}\bigg) \nonumber \\
&\equiv& \frac{c^n_{\rm eff} c_{\rm DM} m_N m_{\rm DM}}{4m_G^2 \Lambda^2}. \label{fn}
\eea
Here, ${\rm DM} = \chi, S, X$ corresponds to fermion, scalar, and vector dark matter, respectively. The effective couplings of quarks ($c_q$) and gluons ($c_g$) to the mediator depend on the interaction structure in the GMDM framework. The scalar form factors for protons ($f^p_{Tq}$) and neutrons ($f^n_{Tq}$) describe the contributions of light quarks ($q = u, d, s$) to the nucleon mass~\cite{Aprile:2017iyp, Akerib:2016vxi, Cui:2017nnn}, while the gluon contribution, $f_{TG}$, accounts for the fraction of the nucleon mass arising from the trace anomaly. Moreover, $q(2)$, $q(2)$ and $G(2)$ are the second moments of the parton distribution functions of quark, antiquark and gluon, respectively.

\begin{itemize}
\item The effective couplings of quarks and gluons to the mediator, $c_q$ and $c_g$, depend on the interaction structure in the GMDM framework. For quarks, $c_q$ is proportional to the overlap of the mediator wavefunction with the quark wavefunction in the extra dimension. For gluons, $c_g$ arises through higher-order loop contributions and the trace anomaly.
\item The scalar form factors for protons ($f^p_{Tq}$) and neutrons ($f^n_{Tq}$) describe the contribution of light quarks ($q = u, d, s$) to the nucleon mass:
\bea
f^p_{Tq} &=& \frac{\langle p | m_q \bar{q} q | p \rangle}{m_p}, \quad f^n_{Tq} = \frac{\langle n | m_q \bar{q} q | n \rangle}{m_n},
\eea
where $m_p$ and $m_n$ are the proton and neutron masses, respectively. Experimental and lattice QCD studies provide estimates for these factors.
\item The gluon contribution, $f_{TG}$, accounts for the fraction of the nucleon mass arising from the gluon field strength via the trace anomaly:
\bea
f_{TG} = 1 - \sum_{q=u,d,s} f^p_{Tq},
\eea
for protons, with an analogous expression for neutrons.
\end{itemize}

Note that this analysis includes the twist-2 gluon operator at tree level, as well as loop effects from heavy quarks and gluons in the trace part, ensuring an accurate prediction of the cross-section.

Direct detection experiments, including XENONnT \cite{XENON:2023cxc},PandaX-4T \cite{PandaX:2024qfu}, XENON1T~\cite{XENON:2018voc,Aprile:2017iyp}, LUX~\cite{Akerib:2016vxi}, and PandaX-II~\cite{Cui:2017nnn}, etc, provide stringent constraints on $\sigma_{{\rm DM}-A}^{SI}$. These constraints translate into bounds on $c_{\rm DM}$, $c_q$, $c_g$, and $\Lambda$, offering a critical test of the GMDM framework. In this work, we impose the most strong constraint from the LZ experiment \cite{LZCollaboration:2024lux} for direct detection.

\section{Collider constraints}
\label{sec:coll}

 The phenomenology of spin-two mediators  depends on the types of couplings they have to Standard Model (SM) particles. As discussed in Sec.\ref{sec:GMDM}, in the bulk benchmark, the couplings exhibit a hierarchy given by:
\begin{equation}
    c_{DM}, c_{H}, c_{t} \gg c_{g,\gamma} \gg c_{f},
\end{equation}
where $c_f$ refers to the couplings to light fermions. By contrast, in the TeV-brane benchmark, all couplings are uniform.

To illustrate the collider reach, we consider two benchmark graviton masses: a relatively light graviton with $m_G = 300$ GeV, and a significantly heavier case with $m_G = 5$ TeV. These choices span a wide range of mediator masses, highlighting both the low- and high-energy regimes relevant for experimental sensitivity.

The production cross section will depend on whether the graviton couples to light quarks, as in the universal benchmark, or not. In the bulk benchmark, the main production mechanism is through gluon fusion. We can then compute the cross section at 13 TeV, the energy of the Run2 LHC whose data we will use to constrain the model. These productions cross sections scale as,
\begin{eqnarray}
   \sigma (q \bar q \to G) &=& \frac{c_q^2}{\Lambda^2 \textrm{ (TeV)}} \, 115 \, (0.022) \textrm{ pb, for $m_G$= 300 GeV (5 TeV)} \nonumber \\
   \sigma (g g \to G) &=& \frac{c_g^2}{\Lambda^2 \textrm{ (TeV)}} \, 4290 \, (0.017) \textrm{ pb, for $m_G$= 300 GeV (5 TeV).}
\end{eqnarray}

\subsection{Limits for the benchmark with universal couplings} 

In the TeV-brane benchmark, the dominant production mechanism for the spin-two mediator is through quark and gluon fusion, with the simplest detectable final state being dileptons. For this scenario, the primary collider sensitivity comes from the process:
\begin{equation}
    \sigma (p \, p \to G \to \ell^+ \ell^-) = \sigma_{\text{prod}} (p \, p \to G ) \times \text{BR}(G\to \ell^+ \ell^-) \times \text{Acc}(m_G),
\end{equation}
where $\text{Acc}(m_G)$ represents the acceptance of the analysis cuts, which depend on the kinematics of the dilepton final state. The cross section is proportional to:
\begin{equation}
    \left(\frac{c_{q} c_\ell}{\Lambda^2}\right)^2,
\end{equation}
where $c_q$ and $c_\ell$ are the couplings of the mediator to quarks and leptons, respectively, and $\Lambda$ is the scale of the effective interaction. The overall production cross section will be the sum of the quark- and gluon-initiated graviton diagrams. Denoting with $c$ the universal coupling, $\sigma(p p \to G)= \frac{c^2}{\Lambda^2}\, 4405 \, (0.04)$ pb for $m_G$= 300 GeV (5 TeV).

For a relatively heavy graviton, the final state leptons are expected to be back-to-back and highly energetic, making them relatively easy to identify. However, as the graviton mass increases further, as in the 5 TeV benchmark, charge identification may become challenging. For example, a highly energetic muon may not curve sufficiently in the magnetic field to allow its charge to be reliably determined. Nonetheless, even in such cases, the event can still be triggered and the invariant mass of the system reconstructed, enabling the identification of the heavy graviton signal. Therefore, for heavy gravitons, the acceptance of the dilepton channel is expected to be of order one~\cite{dileptonsAdam}.

Given this high acceptance, the analysis focuses on the dependence of the cross section on the production cross section and the branching ratio. Both are controlled by the rescaling factors of the couplings and the mediator mass. This approach allows us to explore the collider sensitivity to the TeV-brane benchmark in terms of these parameters, offering insights into the detection prospects of the spin-two mediator.

Searches for resonances decaying into two opposite-sign lepton with the full LHC Run2 dataset\cite{ATLAS:2019erb} place the following 95\% Cl.L. limits in the light and heavy graviton benchmarks,
\begin{eqnarray}
    \sigma(p \, p \to G) \times  BR(G\to \ell^+ \ell^-)   \lesssim  && 2 \textrm{ fb, for } m_X= 300 \textrm{ GeV, and} \nonumber \\
    && 2\times 10^{-2} \textrm{ fb, for } m_X= 5 \textrm{ TeV.}
\end{eqnarray}

\subsection{Limits in the bulk benchmark}

In the bulk benchmark, the couplings of the graviton $G$ to all light fermions are highly suppressed, making gluon fusion the dominant production mechanism. 
Among the possible final states, we consider pairs of gluons, photons, tops, and dibosons ($W^+W^-$, $HH$, $ZZ$). Additionally, the graviton could decay into dark sector particles, $G \to X\, X$, resulting in monojet plus missing energy final states.

For a light graviton with $m_G = 300$ GeV, the phenomenology is dominated by diboson searches. At this mass, the graviton is below the di-top threshold and below the typical selection cut for dijets (typically $m_{jj} > 1$ TeV). While monojet and di-Higgs analyses could also probe this scenario, diboson and diphoton searches provide greater sensitivity. Given that the graviton's couplings to $W$ and $Z$ bosons are larger than its coupling to photons, diboson searches are the most effective channel for setting limits in this low-mass scenario. In Ref.\cite{ATLAS:2020fry}, a search for a so-called bulk graviton was done in both the gluon-fusion (ggF) and vector boson fusion channels (VBF), leading to a limit at 95\% C.L. of
\begin{equation}
    \sigma(p \, p \to G)\times BR(G \to V \, V)   \lesssim   400 \textrm{ fb, for } m_X= 300 \textrm{ GeV,}
\end{equation}
where the VBF channel is the one providing the stringest bound in this case. Other diboson channels, like di-Higgs~\cite{ATLAS:2022hwc}, could become competitive with Run3 data.

For a heavier graviton with $m_G = 5$ TeV, the $t\bar{t}$ final state becomes kinematically accessible, and the dijet final state surpasses the initial selection cut on $m_{jj}$. In this high-mass regime, the dijet channel emerges as the most sensitive probe, benefiting from a straightforward final state and robust acceptance, even as the events become increasingly energetic. In contrast, other channels suffer from a degradation of acceptance due to the highly collimated nature of final states, which challenges isolation criteria. Therefore, in this high-mass scenario, we focus primarily on the dijet final state to assess collider constraints. The current limits from the full Run2 LHC dataset\cite{CMS:2019gwf} are of the order of 
\begin{equation}
   \sigma(p \, p \to G) \times  BR(G\to j \, j)   \lesssim   6 \textrm{ fb, for } m_X= 5 \textrm{ TeV,}
\end{equation}
with X denoting a gluon-gluon resonance, as it is the case for the bulk graviton in our benchmark.

\section{Results}\label{sec:results}


In this section, we present the results of our analysis, focusing on three pairs of plots that illustrate the interplay between relic abundance, direct detection constraints, and collider constraints for different scenarios. These plots demonstrate the viability of the Gravity-Mediated Dark Matter (GMDM) model across various parameter spaces, highlighting both its strengths and its differences from the traditional thermal WIMP paradigm.

\begin{figure}[t!]
\centering
\includegraphics[width=0.43\textwidth,clip]{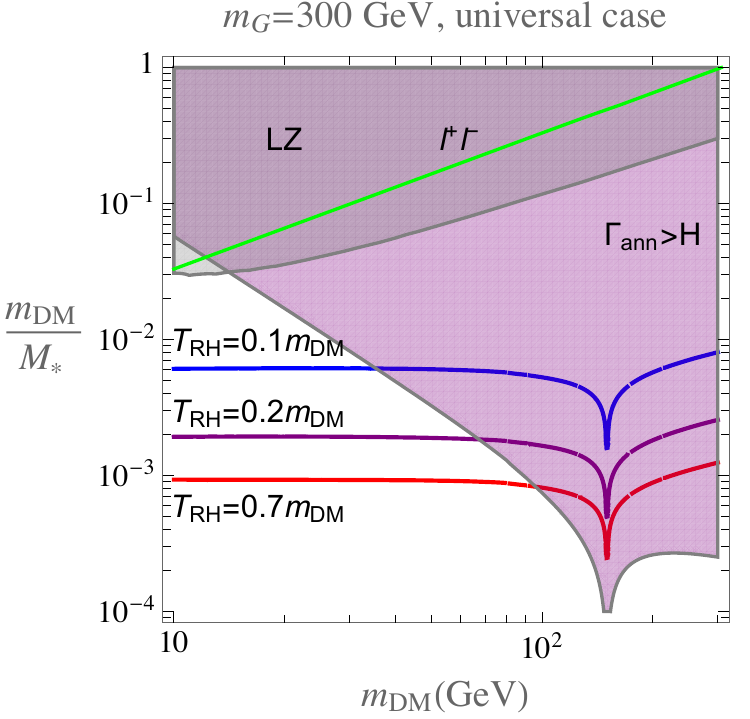}\,\,\,\,\,\,
\includegraphics[width=0.43\textwidth,clip]{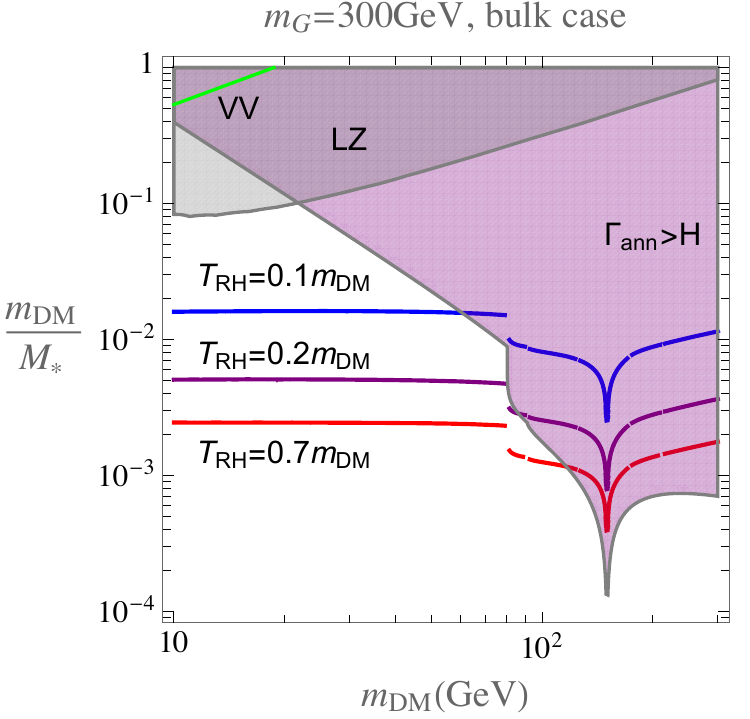}
\caption{Parameter space for  the correct relic density with weak-scale massive graviton and vector dark matter. 
$\Lambda$ denotes the effective cutoff scale, given by $M_*\equiv \sqrt{\Lambda m_G}$. We took $m_G=300\,{\rm GeV}$ and the universal and bulk cases are taken on left and right plots, respectively. The gray region is excluded by LZ experiment with DM-nucleus elastic scattering \cite{LZ}, and the regions above the green lines are excluded by the $l^+l^-$ (diboson) searches at the LHC in the left (right) plot. The DM annihilation rate is larger than the Hubble expansion rate in purple, so the DM couplings required for the correct relic density are subject to small corrections, as compared to those in the approximate results.
}
\label{fig:comparison1}
\end{figure}

{\bf Light Graviton Scenario: }The first pair of plots, Fig.~\ref{fig:comparison1}, corresponds to the light graviton scenario ($m_G = 300$ GeV) and compares the two coupling benchmarks: the universal coupling case and the bulk benchmark. These plots show the ratio $m_{DM}/\Lambda$ on as a function of the mass of the dark matter particle ($m_{DM}$). 

The allowed region for the correct relic abundance is represented as a shaded band, while direct detection constraints exclude certain regions of parameter space (shown in gray). Collider constraints are overlaid as exclusion contours based on current limits (in green). For the bulk benchmark, the hierarchy of couplings reduces direct detection sensitivity, opening a larger viable region compared to the universal coupling case. Both plots demonstrate that the light graviton scenario allows for a substantial region of parameter space where all constraints are satisfied. 

\begin{figure}[t!]
\centering
\includegraphics[width=0.43\textwidth,clip]{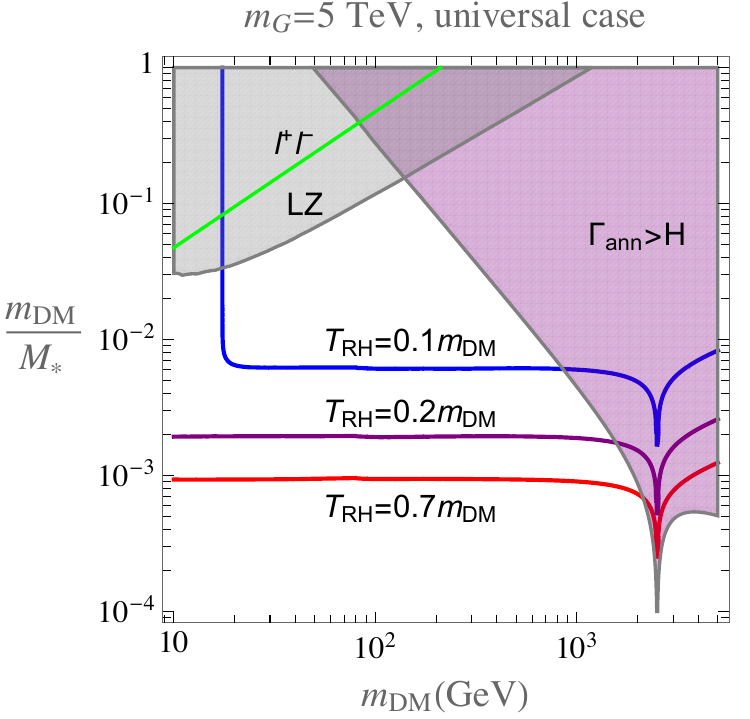}\,\,\,\,\,\,
\includegraphics[width=0.43\textwidth,clip]{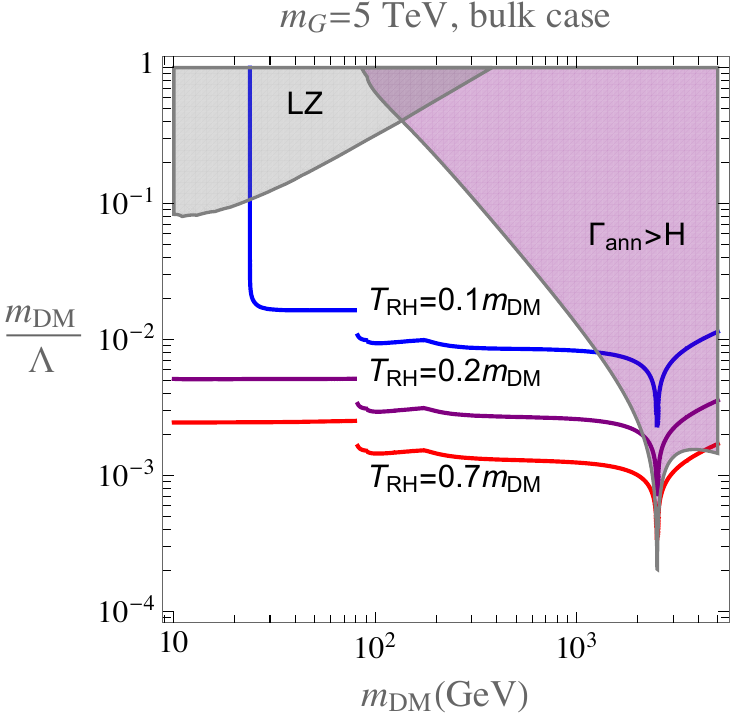}
\caption{Parameter space for the correct relic density with heavy graviton and vector dark matter. We took $m_G=5\,{\rm TeV}$. The color codes are the same as in Fig.~\ref{fig:comparison1}, except that the region above the green line in the left plot is excluded by the $l^+l^-$ searches at the LHC.
}
\label{fig:comparison2}
\end{figure}

{\bf Heavy Graviton Scenario:} The second pair of plots focuses on the heavy graviton scenario ($m_G = 5$ TeV) for the same coupling benchmarks, see Fig.~\ref{fig:comparison2}. These plots again feature  the ratio $m_{DM}/\Lambda$ on as a function of the mass of the dark matter particle ($m_{DM}$), with the same regions for relic abundance, direct detection constraints, and collider constraints. 

In both Figs.~\ref{fig:comparison1} and ~\ref{fig:comparison2}, it is remarkable that the purple regions show that the DM annihilation rate is larger than the Hubble expansion rate near the resonance with the massive graviton, so the backreaction effect due to the DM annihilation cannot be ignored. Even in this case, the corrections due to the DM annihilations to the relic density are mild \cite{lowtemp,bernal}, and the bulk region with relatively light dark matter is still viable.

In this scenario, the direct detection limits are weaker due to the suppression of mediator interactions at low momentum transfer, particularly in the bulk benchmark. Collider constraints, dominated by the dijet channel, play a more significant role in constraining the parameter space. The results indicate that a large viable region remains where relic abundance, direct detection, and collider constraints are simultaneously satisfied.

\begin{figure}[t!]
\centering
\includegraphics[width=0.43\textwidth,clip]{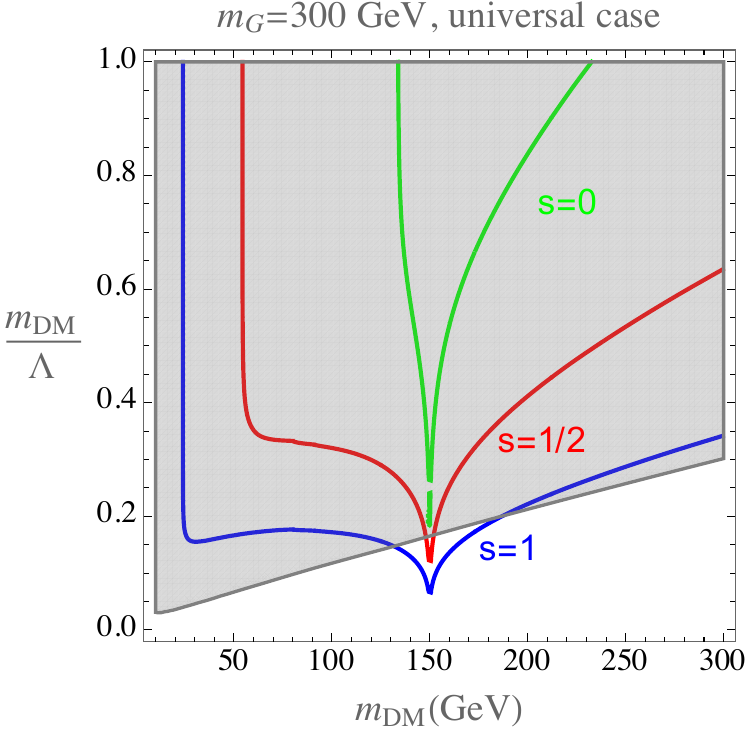}\,\,\,\,\,\,
\includegraphics[width=0.43\textwidth,clip]{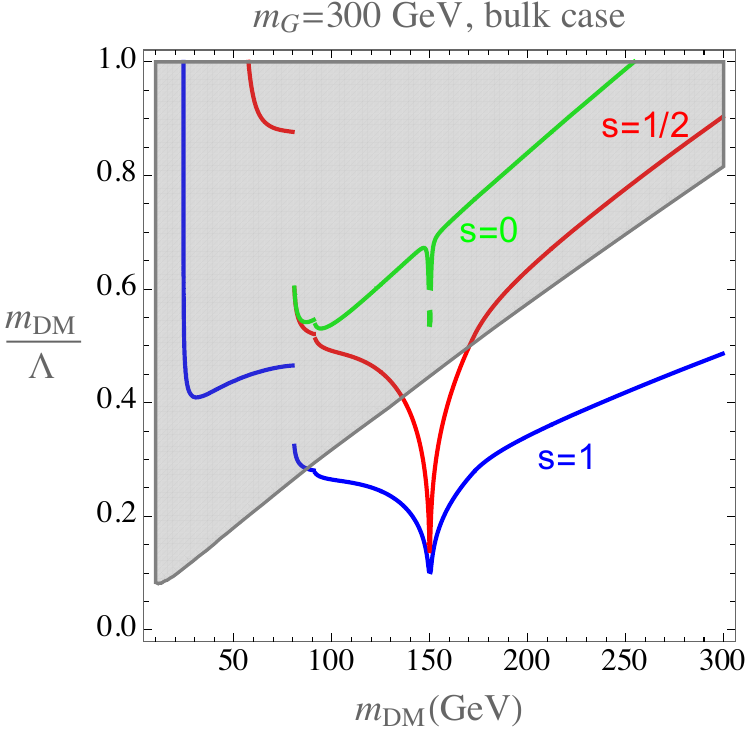}
\caption{Parameter space for  the correct relic density with WIMP dark matter of different spins.
$\Lambda$ denotes the effective cutoff scale, given by $\Lambda\equiv \sqrt{M_* m_G}$. The gray region is excluded by LZ experiment with DM-nucleus elastic scattering \cite{LZ}.
}
\label{fig:wimp}
\end{figure}

{\bf Comparison with the Thermal WIMP Scenario:} The third pair of plots, Fig.~\ref{fig:wimp}, provides a  comparison between the non-thermal GMDM scenarios and the thermal WIMP paradigm which was studied by us in the GMDM case~\cite{GMDM}. For the thermal WIMP case, the relic abundance constraint conflicts significantly with direct detection limits from LZ, leading to the exclusion of most of the parameter space in both universal and bulk benchmarks. In contrast, as we have seen in Figs.~\ref{fig:comparison1} and ~\ref{fig:comparison2}, both the light and heavy graviton non-thermal scenarios exhibit large allowed regions, showcasing the strength of the GMDM framework in evading these tensions.

These results demonstrate that the GMDM paradigm, whether with light or heavy gravitons, offers a compelling alternative to the thermal WIMP scenario. The large parameter space allowed by all constraints highlights the robustness of this framework, particularly in the bulk benchmark, where collider constraints and relic abundance are naturally satisfied with minimal tension from direct detection limits.

\section{Conclusions}
\label{sec:concl}
In this work, we have analyzed the Gravity-Mediated Dark Matter (GMDM) scenario, focusing on the interactions between spin-two mediators, the Standard Model (SM), and dark matter (DM) in a non-thermal production framework. We considered both light and heavy graviton benchmarks, as well as two distinct coupling structures: the universal and bulk scenarios. Our study examined the interplay between relic abundance, direct detection, and collider constraints, providing a comprehensive overview of the parameter space.

Our results demonstrate that both the light graviton ($m_G = 300$ GeV) and heavy graviton ($m_G = 5$ TeV) scenarios offer substantial regions of parameter space where the correct relic abundance is achieved while satisfying current experimental constraints. In the bulk benchmark, where couplings to light fermions are suppressed, direct detection constraints are less stringent, allowing for a larger viable parameter space compared to the universal benchmark.

The collider sensitivity varies significantly with the graviton mass. For light gravitons, diboson and diphoton searches provide the most stringent constraints, while for heavy gravitons, dijet searches dominate. Despite these constraints, viable parameter regions persist in both cases, demonstrating the robustness of the GMDM framework.

A comparison with the thermal WIMP paradigm shows the advantages of the non-thermal production mechanisms explored in this study. While the thermal WIMP scenario is excluded due to conflicting relic abundance and direct detection constraints, the non-thermal GMDM framework avoids these issues and remains a viable explanation for dark matter.

This work highlights the potential of GMDM as an alternative to the thermal relic paradigm, offering greater flexibility in accommodating experimental data. The non-thermal framework expands the scope of dark matter model building by relaxing the constraints imposed by thermal equilibrium. Future studies could extend this analysis to explore other production mechanisms, mediator mass ranges, and coupling structures, as well as incorporate refined experimental constraints as new data becomes available. 

\section*{Acknowledgments}

HML is supported in part by Basic Science Research Program through the National
Research Foundation of Korea (NRF) funded by the Ministry of Education, Science and
Technology (NRF-2022R1A2C2003567). 
The research of VS is supported by the Generalitat
Valenciana PROMETEO/2021/083, Proyecto Consolidacion CNS2022-135688,  the Ministerio de Ciencia e
Innovacion projects PID2020-113644GB-I00 and PID2023-148162NB-C21, and the {\it Severo Ochoa} project CEX2023-001292-S funded by MCIU/AEI.

\def\theequation{A.\arabic{equation}}

\setcounter{equation}{0}
\vskip0.8cm
\noindent

\appendix
\section{DM annihilation cross sections}\label{appA}

We list the annihilation cross sections for non-relativistic dark matter with different spins before thermal average.
Parametrizing the leading annihilation cross section by $(\sigma v)=\sigma_n v^{2n}$, where $n=0, 1, 2$ for s-wave, p-wave and d-wave, respectively,  we obtain the thermal averaged cross section as $\langle\sigma v\rangle= 2^n (2n+1)!! \sigma_n x^{-n}$.  So, we get $\langle\sigma v\rangle=\sigma_0$, $6\sigma_1 x^{-1}$ and $60\sigma_2 x^{-2}$, for s-wave, p-wave and d-wave, respectively,

{\bf Scalar dark matter}:

The annihilation cross section for scalar dark matter into a pair of SM fermions, $SS\rightarrow \psi{\bar\psi}$, is given \cite{GMDM} by 
\bea
(\sigma v)_{SS\rightarrow\psi{\bar\psi} } = v^4 \cdot  \frac{ N_c c_S^2 c_\psi^2 }{360\pi \Lambda^4} 
\frac{m_S^6}{(m_G^2-4 m_S^2)^2}
\left(1-\frac{m_\psi^2}{m_S^2}\right)^\frac{3}{2} \left(3+\frac{2m_\psi^2}{m_S^2}\right)
\eea
where $N_c$ is the number of colors for the SM fermion $\psi$.

For $m_S>m_G$, scalar dark matter can also annihilate into a pair of spin-2 particles through the $t/u$-channels \cite{GMDM,GMDM-dd}, so the corresponding annihilation cross section is given, as follows,
\bea
(\sigma v)_{SS\rightarrow GG} = \frac{4 c_{S}^4 m_{S}^2}{9 \pi \Lambda^4 }
\frac{(1-r_S)^\frac{9}{2}}{r^4_S  (2-r_S)^2}   \label{tch-scalar}
\eea
with $r_S = \left(\frac{m_G}{m_S}\right)^2$.

The annihilation cross sections into a pair of massless gauge bosons \cite{GMDM} are
\bea
(\sigma v)_{SS\rightarrow \gamma\gamma}&=&v^4 \cdot  \frac{c^2_S c^2_\gamma }{60\pi\Lambda^4}\frac{m^6_S}{(4m^2_S-m^2_G)^2+\Gamma^2_G m^2_G},\\
(\sigma v)_{SS\rightarrow gg}&=& v^4 \cdot  \frac{2c^2_S c^2_g }{15\pi\Lambda^4}\frac{m^6_S}{(4m^2_S-m^2_G)^2+\Gamma^2_G m^2_G}.
\eea

The annihilation cross section of scalar dark matter annihilating into a pair of Higgs bosons \cite{GMDM} is
\bea
(\sigma v)_{SS\rightarrow hh} \simeq v^4 \cdot\frac{ c^2_S c^2_H}{720\pi \Lambda^4}\, \frac{m^6_S}{(4m^2_S-m^2_G)^2+\Gamma^2_G m^2_G} \left(1-\frac{m^2_h}{m^2_S}\right)^\frac{5}{2}.
\eea
The  annihilation cross sections of scalar dark matter into a pair of massive gauge bosons are
\bea
(\sigma v)_{SS\rightarrow Z Z}&=& \frac{3c^2_S(c_V-c_H)^2}{16\pi \Lambda^4} \frac{m^2_S m^4_Z}{(4m^2_S-m^2_G)^2+\Gamma^2_G m^2_G}\left(1-\frac{4m^2_S}{m^2_G}\right)^2\left(1-\frac{m^2_Z}{m^2_S}\right)^{\frac{1}{2}},  \\
(\sigma v)_{SS\rightarrow W W}&=& \frac{3c^2_S(c_V-c_H)^2}{8\pi \Lambda^4} \frac{m^2_S m^4_W}{(4m^2_S-m^2_G)^2+\Gamma^2_G m^2_G}\left(1-\frac{4m^2_S}{m^2_G}\right)^2\left(1-\frac{m^2_Z}{m^2_S}\right)^{\frac{1}{2}}.
\eea
For $c_H=c_V$, both s-wave and p-wave components are zero and the annihilation cross section becomes d-wave as
\bea
(\sigma v)_{SS\rightarrow Z Z}&=&  v^4 \cdot 
\frac{c^2_S c^2_V}{720\pi \Lambda^4} \frac{m^6_S}{(4m^2_S-m^2_G)^2+\Gamma^2_G m^2_G}\left(1-\frac{m^2_Z}{m^2_S}\right)^{\frac{1}{2}}
\left(13+\frac{14m^2_Z}{m^2_S}+\frac{3m^4_Z}{m^4_S}\right), \\
(\sigma v)_{SS\rightarrow WW}&=&  v^4 \cdot
\frac{c^2_S c^2_V }{360\pi \Lambda^4} \frac{m^6_S}{(4m^2_S-m^2_G)^2+\Gamma^2_G m^2_G}\left(1-\frac{m^2_W}{m^2_S}\right)^\frac{1}{2}
\left(13+\frac{14m^2_W}{m^2_S}+\frac{3m^4_W}{m^4_S}\right). \quad
\eea

{\bf Fermion dark matter}:

The annihilation cross section for fermion dark matter, $\chi{\bar\chi}\rightarrow \psi{\bar\psi}$, is given \cite{GMDM} by 
\bea
(\sigma v)_{\chi{\bar\chi}\rightarrow \psi{\bar\psi}} = v^2 \cdot \frac{N_c c^2_\chi c^2_\psi }{72\pi\Lambda^4}
\frac{m^6_\chi}{(4m^2_\chi-m^2_G)^2+\Gamma^2_G m^2_G} \left(1-\frac{m^2_\psi}{m^2_\chi}\right)^\frac{3}{2} 
\left(3+\frac{2m^2_\psi}{m^2_\chi}\right).
\eea

For $m_{\chi}>m_G$, fermion dark matter also annihilates into a pair of spin-2 particles through  to the $t/u$-channels \cite{GMDM,GMDM-dd}, as follows,
\bea
(\sigma v)_{\chi \bar\chi \rightarrow GG} &=& \frac{c_{\chi}^4 m_{\chi}^2}{16 \pi \Lambda^4 }
\frac{(1-r_\chi)^\frac{7}{2}}{r^2_\chi (2-r_\chi)^2}  \label{tch-fermion}
\eea
with $r_\chi = \left(\frac{m_G}{m_\chi}\right)^2$.

The annihilation cross sections into a pair of massless gauge bosons and a pair of mesons \cite{GMDM} are
\bea
(\sigma v)_{\chi{\bar\chi}\rightarrow \gamma\gamma}&=& v^2\cdot \frac{c^2_\chi c^2_\gamma}{12\pi\Lambda^4}\frac{m^6_\chi}{(4m^2_\chi-m^2_G)^2+\Gamma^2_G m^2_G},\\
(\sigma v)_{\chi{\bar\chi}\rightarrow gg}&=&  v^2\cdot\frac{2c^2_\chi c^2_g}{3\pi\Lambda^4}\frac{m^6_\chi}{(4m^2_\chi-m^2_G)^2+\Gamma^2_G m^2_G}.
\eea

The annihilation cross section of fermion dark matter annihilating to a pair of Higgs bosons is
\bea
(\sigma v)_{\chi{\bar\chi}\rightarrow hh}= v^2\cdot \frac{c^2_\chi c^2_H }{144\pi\Lambda^4}
\frac{m^6_\chi}{(4m^2_\chi-m^2_G)^2+\Gamma^2_G m^2_G} \left(1-\frac{m^2_h}{m^2_\chi}\right)^\frac{5}{2}.
\eea
The annihilation cross sections for fermion dark matter going into a pair of massive gauge bosons are
\bea
(\sigma v)_{\chi{\bar\chi}\rightarrow ZZ} &=&  v^2 \cdot \frac{c_\chi^2 c_V^2}{144\pi \Lambda^4} 
\frac{m_\chi^6}{(m_G^2-4 m_\chi^2)^2+\Gamma_G^2 m_G^2} \left(1-\frac{m_Z^2}{m_\chi^2}\right)^\frac{1}{2}  \nonumber \\
&&\times \Bigg[\left(13+\frac{14m_Z^2}{m_\chi^2}+\frac{3m_Z^4}{m_\chi^4}\right)-2  \left(1-\frac{c_H}{c_V}\right) \left(1+\frac{13m_Z^2}{m_\chi^2}+\frac{m_Z^4}{m_\chi^4}\right)\nonumber \\
&&\quad+ \left(1-\frac{c_H}{c_V}\right)^2 \left\{1+\frac{3m_Z^2}{m_\chi^2}+\frac{31}{8}\frac{m_Z^4}{m_\chi^4}
-\frac{3m_Z^4}{m_G^2 m_\chi^2}+\frac{6m_Z^4}{m_G^4}
\right\}
\Bigg] , \\
(\sigma v)_{\chi{\bar\chi}\rightarrow WW} &=&  v^2 \cdot \frac{c_\chi^2 c_V^2}{72\pi \Lambda^4} 
\frac{m_\chi^6}{(m_G^2-4 m_\chi^2)^2+\Gamma_G^2 m_G^2} \left(1-\frac{m_W^2}{m_\chi^2}\right)^\frac{1}{2}  \nonumber \\
&&\times  \Bigg[\left(13+\frac{14m_W^2}{m_\chi^2}+\frac{3m_W^4}{m_\chi^4}\right)-2  \left(1-\frac{c_H}{c_V}\right) \left(1+\frac{13m_W^2}{m_\chi^2}+\frac{m_W^4}{m_\chi^4}\right)\nonumber \\
&&\quad+\left(1-\frac{c_H}{c_V}\right)^2 \left\{1+\frac{3m_W^2}{m_\chi^2}+\frac{31}{8}\frac{m_W^4}{m_\chi^4}
-\frac{3m_W^4}{m_G^2 m_\chi^2}+\frac{6m_W^4}{m_G^4}
\right\}
\Bigg].
\eea
For $c_H=c_V$, the above annihilation cross sections become
\bea
(\sigma v)_{\chi{\bar\chi}\rightarrow ZZ}&=& v^2\cdot\frac{c^2_\chi c^2_V}{144\pi\Lambda^4}\frac{m^6_\chi}{(4m^2_\chi-m^2_G)^2+\Gamma^2_G m^2_G} \left(13+\frac{14m^2_Z}{m^2_\chi}+\frac{3m^4_Z}{m^4_\chi}\right)\left(1-\frac{m^2_Z}{m^2_\chi}\right)^\frac{1}{2}, \\
(\sigma v)_{\chi{\bar\chi}\rightarrow WW}&= & v^2\cdot \frac{c^2_\chi c^2_V}{72\pi\Lambda^4}\frac{m^6_\chi}{(4m^2_\chi-m^2_G)^2+\Gamma^2_G m^2_G}  \left(13+\frac{14m^2_W}{m^2_\chi}+\frac{3m^4_W}{m^4_\chi}\right)\left(1-\frac{m^2_W}{m^2_\chi}\right)^\frac{1}{2}.
\eea

{\bf Vector dark matter}:

The annihilation cross section for vector dark matter, $XX\rightarrow \psi{\bar\psi}$, is given \cite{GMDM} by
\bea
(\sigma v)_{XX\rightarrow \psi{\bar\psi}}&=& \frac{4N_c c^2_X c^2_\psi }{27\pi \Lambda^4}
 \frac{m^6_X}{(4m^2_X-m^2_G)^2+\Gamma^2_G m^2_G}\left(3+\frac{2m^2_\psi}{m^2_X}\right)\left(1-\frac{m^2_\psi}{m^2_X}\right)^\frac{3}{2}.
\eea

For $m_X>m_G$,  vector dark matter also annihilates into a pair of spin-2 particles through  the $t/u$-channels \cite{GMDM,GMDM-dd}, as follows,
\bea
(\sigma v)_{X X  \rightarrow GG} &=&
\frac{c_{X}^4 m_{X}^2}{324 \pi \Lambda^4 }
\frac{\sqrt{1-r_X}}{r^4_X  (2-r_X)^2} \,
\bigg(176+192 r_X+1404 r^2_X-3108 r^3_X \nonumber \\
&&+1105 r^4_X+362 r^5_X+34 r^6_X \bigg) 
\label{tch-vector}
\eea
with $r_X = \left(\frac{m_G}{m_X}\right)^2$.

The annihilation cross sections into a pair of massless gauge bosons and a pair of mesons \cite{GMDM} are
\bea
(\sigma v)_{XX\rightarrow \gamma\gamma}&=&\frac{8c^2_X c^2_\gamma}{9\pi\Lambda^4}\frac{m^6_X}{(4m^2_X-m^2_G)^2+\Gamma^2_G m^2_G}, \\
(\sigma v)_{XX\rightarrow gg}&=&\frac{64c^2_X c^2_g}{9\pi\Lambda^4}\frac{m^6_X}{(4m^2_X-m^2_G)^2+\Gamma^2_G m^2_G}.
\eea

The annihilation cross section of vector dark matter annihilating into a pair of Higgs bosons is
\bea
(\sigma v)_{XX\rightarrow hh}=\frac{2c^2_X c^2_H}{27\pi \Lambda^4}
\frac{m^6_X}{(4m^2_X-m^2_G)^2+\Gamma^2_G m^2_G}\left(1-\frac{m^2_h}{m^2_X}\right)^\frac{5}{2}.
\eea
The annihilation cross sections for vector dark matter going into a pair of massive gauge bosons are
\bea
(\sigma v)_{XX\rightarrow ZZ} &=&  \frac{2 c_{X}^2 c_V^2}{27\pi \Lambda^4} 
\frac{m_{X}^6}{(m_G^2-4 m_{X}^2)^2+\Gamma_G^2 m_G^2} \left(1-\frac{m_Z^2}{m_{X}^2}\right)^\frac{1}{2} \nonumber \\
&&\times \Bigg[\left(13+\frac{14m_Z^2}{m_{X}^2}+\frac{3m_Z^4}{m_{X}^4}\right)
-2  \left(1-\frac{c_H}{c_V}\right) \left(1+\frac{13m_Z^2}{m_{X}^2}+\frac{m_Z^4}{m_{X}^4}\right)\nonumber \\
&&\quad + \left(1-\frac{c_H}{c_V}\right)^2 \left\{1+\frac{3m_Z^2}{m_{X}^2}+\frac{115}{32}\frac{m_Z^4}{m_{X}^4}
-\frac{3}{4}\frac{m_Z^4}{m_G^2 m_{X}^2}+\frac{3}{2}\frac{m_Z^4}{m_G^4}
\right\}
\Bigg]  , \\
(\sigma v)_{XX\rightarrow WW} &=&  \frac{4 c_{X}^2 c_V^2}{27\pi \Lambda^4} 
\frac{m_{X}^6}{(m_G^2-4 m_{X}^2)^2+\Gamma_G^2 m_G^2}  \left(1-\frac{m_W^2}{m_{X}^2}\right)^\frac{1}{2} \nonumber \\
&&\times \Bigg[\left(13+\frac{14m_W^2}{m_{X}^2}+\frac{3m_W^4}{m_{X}^4}\right)
-2  \left(1-\frac{c_H}{c_V}\right) \left(1+\frac{13m_W^2}{m_{X}^2}+\frac{m_W^4}{m_{X}^4}\right)\nonumber \\
&&\quad+ \left(1-\frac{c_H}{c_V}\right)^2 \left\{1+\frac{3m_W^2}{m_{X}^2}+\frac{115}{32}\frac{m_W^4}{m_{X}^4}
-\frac{3}{4}\frac{m_W^4}{m_G^2 m_{X}^2}+\frac{3}{2}\frac{m_W^4}{m_G^4}
\right\}
\Bigg] .
\eea 
For $c_H=c_V$, the above annihilation cross sections become
\bea
(\sigma v)_{XX\rightarrow ZZ} &=&  \frac{2 c_{X}^2 c_V^2}{27\pi \Lambda^4} 
\frac{m_{X}^6}{(m_G^2-4 m_{X}^2)^2+\Gamma_G^2 m_G^2} 
\left(13+\frac{14m_Z^2}{m_{X}^2}+\frac{3m_Z^4}{m_{X}^4}\right)
 \left(1-\frac{m_Z^2}{m_{X}^2}\right)^\frac{1}{2} , \\
(\sigma v)_{XX\rightarrow WW} &=&  \frac{4 c_{X}^2 c_V^2}{27\pi \Lambda^4} 
\frac{m_{X}^6}{(m_G^2-4 m_{X}^2)^2+\Gamma_G^2 m_G^2} 
\left(13+\frac{14m_W^2}{m_{X}^2}+\frac{3m_W^4}{m_{X}^4}\right) \left(1-\frac{m_W^2}{m_{X}^2}\right)^\frac{1}{2}  .
\eea 

\section{Graviton's partial widths}
\label{appB}

In this appendix, we show the analytical expressions of the KK graviton decay rates, as were discussed in Ref.~\cite{GMDM}.

The partial width to two Higgs particles is given by,
\bea
\Gamma(G \rightarrow hh)= \frac{c^2_H m^3_G}{960 \pi \Lambda^2} \Big(1-\frac{4m^2_h}{m^2_G}\Big)^\frac{5}{2}.\label{Ghh} \,
\eea
whereas the width to two massive vector bosons receives contributions from the coupling to the gauge field $c_V$ and to the Higgs doublet $c_H$ where both transverse and longitudinal degrees of freedom can contribute, 
\bea\label{ec:GaVV}
\Gamma(G \rightarrow ZZ)&=& \frac{m^3_G}{960 \pi \Lambda^2}
\bigg[c^2_H\Big(1+\frac{12m^2_Z}{m^2_G}+\frac{56m^4_Z}{m^4_G}\Big)+80 c_V c_H \Big(1-\frac{m^2_Z}{m^2_G}\Big)\frac{m^2_Z}{m^2_G} \nonumber \\
&&+12c^2_V\Big(1-\frac{3m^2_Z}{m^2_G}+\frac{6m^4_Z}{m^4_G}\Big)  \bigg]\Big(1-\frac{4m^2_Z}{m^2_G}\Big)^\frac{1}{2}, \\
\Gamma(G \rightarrow WW)&=&\frac{m^3_G}{480 \pi \Lambda^2}
\bigg[c^2_H\Big(1+\frac{12m^2_W}{m^2_G}+\frac{56m^4_W}{m^4_G}\Big)+80 c_V c_H \Big(1-\frac{m^2_W}{m^2_G}\Big)\frac{m^2_W}{m^2_G} \nonumber \\
&&+12c^2_V\Big(1-\frac{3m^2_W}{m^2_G}+\frac{6m^4_W}{m^4_G}\Big)  \bigg]\Big(1-\frac{4m^2_W}{m^2_G}\Big)^\frac{1}{2}.
\eea

When $c_V=c_H$ (Universal Benchmark), the decay rates into a pair of massive gauge bosons simplify to
\bea
\Gamma(G \rightarrow Z Z)&=& \frac{c^2_V m^3_G}{960\pi \Lambda^2}\Big(1- \frac{4m^2_Z}{m^2_G}\Big)^\frac{1}{2}
\Big(13+\frac{56m^2_Z}{m^2_G}+\frac{48m^4_Z}{m^4_G}\Big), \\
\Gamma(G \rightarrow W W )&=& \frac{c^2_V m^3_G}{480\pi \Lambda^2}\Big(1- \frac{4m^2_W}{m^2_G}\Big)^\frac{1}{2}
\Big(13+ \frac{56m^2_W}{m^2_G}+\frac{48m^4_W}{m^4_G}\Big).
\eea
The Graviton's width to massless gauge bosons (photons and gluons) is as follows,
\bea\label{ec:GaAA}
\Gamma(G \rightarrow\gamma\gamma)&=& \frac{c^2_\gamma m^3_G}{80\pi\Lambda^2}, \\
\Gamma(G \rightarrow gg)&=&  \frac{c^2_g m^3_G}{10\pi\Lambda^2}.
\eea

And finally, the widths to a pair of DM scalars, fermions or gauge bosons, are given by 
\bea\label{ec:Gaff}
\Gamma(G \rightarrow SS)&=&  \frac{c_S^2 m^3_G}{960 \pi \Lambda^2} \Big(1-\frac{4m^2_S}{m^2_G}\Big)^\frac{5}{2}, \\
\Gamma(G \rightarrow\chi{\bar\chi})&=& \frac{ c_\chi^2 m^3_G}{160 \pi \Lambda^2}
 \left(1-\frac{4m^2_\chi}{m^2_G} \right)^\frac{3}{2} \left(1+\frac{8}{3} \frac{m^2_\chi}{m^2_G}\right), \\
 \Gamma(G \rightarrow XX)&=& \frac{ c_X^2 m^3_G}{960\pi \Lambda^2}\Big(1- \frac{4m^2_X}{m^2_G}\Big)^\frac{1}{2}
\Big(13+\frac{56m^2_X}{m^2_G}+\frac{48m^4_X}{m^4_G}\Big).
\eea


\begin{thebibliography}{999}

\bibitem{lowtemp}
C.~Cosme, F.~Costa and O.~Lebedev,
``Freeze-in at stronger coupling,''
Phys. Rev. D \textbf{109} (2024) no.7, 075038
doi:10.1103/PhysRevD.109.075038
[arXiv:2306.13061 [hep-ph]];
C.~Cosme, F.~Costa and O.~Lebedev,
``Temperature evolution in the Early Universe and freeze-in at stronger coupling,''
JCAP \textbf{06} (2024), 031
doi:10.1088/1475-7516/2024/06/031
[arXiv:2402.04743 [hep-ph]];
G.~Arcadi, F.~Costa, A.~Goudelis and O.~Lebedev,
``Higgs portal dark matter freeze-in at stronger coupling: observational benchmarks,''
JHEP \textbf{07} (2024), 044
doi:10.1007/JHEP07(2024)044
[arXiv:2405.03760 [hep-ph]].

\bibitem{Boddy:2024vgt}
K.~K.~Boddy, K.~Freese, G.~Montefalcone and B.~Shams Es Haghi,
[arXiv:2405.06226 [hep-ph]].


\bibitem{bernal}
N.~Bernal, K.~Deka and M.~Losada,
JCAP \textbf{09} (2024), 024
doi:10.1088/1475-7516/2024/09/024
[arXiv:2406.17039 [hep-ph]].

\bibitem{GMDM}
  H.~M.~Lee, M.~Park and V.~Sanz,
  ``Gravity-mediated (or Composite) Dark Matter,''
  Eur.\ Phys.\ J.\ C {\bf 74} (2014) 2715
  doi:10.1140/epjc/s10052-014-2715-8
  [arXiv:1306.4107 [hep-ph]];
  H.~M.~Lee, M.~Park and V.~Sanz,
  ``Gravity-mediated (or Composite) Dark Matter Confronts Astrophysical Data,''
  JHEP {\bf 1405} (2014) 063
  doi:10.1007/JHEP05(2014)063
  [arXiv:1401.5301 [hep-ph]].

\bibitem{dual}
 V.~Sanz,
  ``On the compatibility of the diboson excess with a gg-initiated composite sector,''
  arXiv:1507.03553 [hep-ph].
 R.~Fok, C.~Guimaraes, R.~Lewis and V.~Sanz,
  ``It is a Graviton! or maybe not,''
  JHEP {\bf 1212} (2012) 062
  doi:10.1007/JHEP12(2012)062
  [arXiv:1203.2917 [hep-ph]].

\bibitem{RS} 
 M.~Gogberashvili,
  ``Hierarchy problem in the shell universe model,''
  Int.\ J.\ Mod.\ Phys.\ D {\bf 11}, 1635 (2002)
  [hep-ph/9812296].
   M.~Gogberashvili,
  ``Our world as an expanding shell,''
  Europhys.\ Lett.\  {\bf 49}, 396 (2000)
  [hep-ph/9812365].
   M.~Gogberashvili,
  ``Four dimensionality in noncompact Kaluza-Klein model,''
  Mod.\ Phys.\ Lett.\ A {\bf 14}, 2025 (1999)
  [hep-ph/9904383].
 L.~Randall and R.~Sundrum,
  ``A Large mass hierarchy from a small extra dimension,''
  Phys.\ Rev.\ Lett.\  {\bf 83}, 3370 (1999)
  [hep-ph/9905221].

\bibitem{KS}
  I.~R.~Klebanov and M.~J.~Strassler,
  ``Supergravity and a confining gauge theory: Duality cascades and chi SB resolution of naked singularities,''
  JHEP {\bf 0008} (2000) 052
  [hep-th/0007191].



\bibitem{localization}
  H.~Davoudiasl, J.~L.~Hewett and T.~G.~Rizzo,
  ``Experimental probes of localized gravity: On and off the wall,''
  Phys.\ Rev.\ D {\bf 63}, 075004 (2001)
  [hep-ph/0006041];
  B.~Batell and T.~Gherghetta,
  ``Localized U(1) gauge fields, millicharged particles, and holography,''
  Phys.\ Rev.\ D {\bf 73}, 045016 (2006)
  [hep-ph/0512356];
  B.~Batell and T.~Gherghetta,
  ``Yang-Mills Localization in Warped Space,''
  Phys.\ Rev.\ D {\bf 75}, 025022 (2007)
  [hep-th/0611305].

\bibitem{CHM}
   F.~Caracciolo, A.~Parolini and M.~Serone,
  ``UV Completions of Composite Higgs Models with Partial Compositeness,''
  JHEP {\bf 1302}, 066 (2013)
  [arXiv:1211.7290 [hep-ph]].
  M.~Redi and A.~Weiler,
  ``Flavor and CP Invariant Composite Higgs Models,''
  JHEP {\bf 1111}, 108 (2011)
  [arXiv:1106.6357 [hep-ph]].
   S.~De Curtis, M.~Redi and A.~Tesi,
  ``The 4D Composite Higgs,''
  JHEP {\bf 1204}, 042 (2012)
  [arXiv:1110.1613 [hep-ph]].
  
 \bibitem{PomarolDM}
 M.~Frigerio, A.~Pomarol, F.~Riva and A.~Urbano,
  ``Composite Scalar Dark Matter,''
  JHEP {\bf 1207}, 015 (2012)
  [arXiv:1204.2808 [hep-ph]].


\bibitem{chala}
 M.~Chala,
  ``$h \rightarrow \gamma\gamma$ excess and Dark Matter from Composite Higgs Models,''
  JHEP {\bf 1301}, 122 (2013)
  [arXiv:1210.6208 [hep-ph]].
  
   
\bibitem{cmbV}
M.~A.~Sanchis-Lozano and V.~Sanz,
Phys. Rev. D \textbf{109} (2024) no.6, 063529
doi:10.1103/PhysRevD.109.063529
[arXiv:2312.02740 [astro-ph.CO]].
\bibitem{RSbulk}
  A.~L.~Fitzpatrick, J.~Kaplan, L.~Randall and L.~-T.~Wang,
  ``Searching for the Kaluza-Klein Graviton in Bulk RS Models,''
  JHEP {\bf 0709} (2007) 013
  [hep-ph/0701150].
T.~Gherghetta and A.~Pomarol,
  ``Bulk fields and supersymmetry in a slice of AdS,''
  Nucl.\ Phys.\ B {\bf 586}, 141 (2000)
  [hep-ph/0003129].
   T.~Gherghetta and A.~Pomarol,
  ``A Warped supersymmetric standard model,''
  Nucl.\ Phys.\ B {\bf 602}, 3 (2001)
  [hep-ph/0012378].
 Y.~Grossman and M.~Neubert,
  ``Neutrino masses and mixings in nonfactorizable geometry,''
  Phys.\ Lett.\ B {\bf 474}, 361 (2000)
  [hep-ph/9912408].
   H.~Davoudiasl, J.~L.~Hewett and T.~G.~Rizzo,
  ``Bulk gauge fields in the Randall-Sundrum model,''
  Phys.\ Lett.\ B {\bf 473} (2000) 43
  [hep-ph/9911262].
    A.~Pomarol,
  ``Gauge bosons in a five-dimensional theory with localized gravity,''
  Phys.\ Lett.\ B {\bf 486} (2000) 153
  [hep-ph/9911294].
    S.~Chang, J.~Hisano, H.~Nakano, N.~Okada and M.~Yamaguchi,
  ``Bulk standard model in the Randall-Sundrum background,''
  Phys.\ Rev.\ D {\bf 62} (2000) 084025
  [hep-ph/9912498].
    G.~F.~Giudice, R.~Rattazzi and J.~D.~Wells,
  ``Graviscalars from higher dimensional metrics and curvature Higgs mixing,''
  Nucl.\ Phys.\ B {\bf 595} (2001) 250
  [hep-ph/0002178].
    H.~Davoudiasl, J.~L.~Hewett and T.~G.~Rizzo,
  ``Experimental probes of localized gravity: On and off the wall,''
  Phys.\ Rev.\ D {\bf 63} (2001) 075004
  [hep-ph/0006041].
   K.~Agashe, H.~Davoudiasl, G.~Perez and A.~Soni,
  ``Warped Gravitons at the LHC and Beyond,''
  Phys.\ Rev.\ D {\bf 76} (2007) 036006
  [hep-ph/0701186].
 R.~Bao, M.~S.~Carena, J.~Lykken, M.~Park and J.~Santiago,
  ``Revamped braneworld gravity,''
  Phys.\ Rev.\ D {\bf 73} (2006) 064026
  [hep-th/0511266].
  
\bibitem{fdual1} 

T.~Gherghetta,
``Les Houches lectures on warped models and holography,''
[arXiv:hep-ph/0601213 [hep-ph]].

 \bibitem{AdSCFTgen}
 See for example the following review, and references therein.
  O.~Aharony, S.~S.~Gubser, J.~M.~Maldacena, H.~Ooguri and Y.~Oz,
  ``Large N field theories, string theory and gravity,''
  Phys.\ Rept.\  {\bf 323}, 183 (2000)
  [hep-th/9905111]. 
  
\bibitem{LisaPoratti} 
  N.~Arkani-Hamed, M.~Porrati and L.~Randall,
  ``Holography and phenomenology,''
  JHEP {\bf 0108}, 017 (2001)
  [hep-th/0012148].


 \bibitem{analogue} 
  J.~Hirn and V.~Sanz,
  ``The Fifth dimension as an analogue computer for strong interactions at the LHC,''
  JHEP {\bf 0703}, 100 (2007)
  [hep-ph/0612239].
   J.~Hirn and V.~Sanz,
  ``A Negative S parameter from holographic technicolor,''
  Phys.\ Rev.\ Lett.\  {\bf 97}, 121803 (2006)
  [hep-ph/0606086].
  J.~Hirn, A.~Martin and V.~Sanz,
  ``Benchmarks for new strong interactions at the LHC,''
  JHEP {\bf 0805}, 084 (2008)
  [arXiv:0712.3783 [hep-ph]].
  
\bibitem{lisa-matt}
 L.~Randall, V.~Sanz and M.~D.~Schwartz,
  ``Entropy area relations in field theory,''
  JHEP {\bf 0206}, 008 (2002)
  [hep-th/0204038].
  
  \bibitem{csaba-fat}
   C.~Csaki, Y.~Shirman and J.~Terning,
  ``A Seiberg Dual for the MSSM: Partially Composite W and Z,''
  Phys.\ Rev.\ D {\bf 84}, 095011 (2011)
  [arXiv:1106.3074 [hep-ph]].
  
\bibitem{ami-qcd}
J.~Erlich, E.~Katz, D.~T.~Son and M.~A.~Stephanov,
  ``QCD and a holographic model of hadrons,''
  Phys.\ Rev.\ Lett.\  {\bf 95}, 261602 (2005)
  [hep-ph/0501128].
  
 \bibitem{alex-qcd} 
   L.~Da Rold and A.~Pomarol,
  ``Chiral symmetry breaking from five dimensional spaces,''
  Nucl.\ Phys.\ B {\bf 721}, 79 (2005)
  [hep-ph/0501218].
  
\bibitem{QCD-hol}
 J.~Hirn and V.~Sanz,
  ``Interpolating between low and high energy QCD via a 5-D Yang-Mills model,''
  JHEP {\bf 0512} (2005) 030
  [hep-ph/0507049].
  J.~Hirn, N.~Rius and V.~Sanz,
  ``Geometric approach to condensates in holographic QCD,''
  Phys.\ Rev.\  D {\bf 73}, 085005 (2006)
  [arXiv:hep-ph/0512240].
  
\bibitem{massives2}
 E.~Kiritsis,
  ``Product CFTs, gravitational cloning, massive gravitons and the space of gravitational duals,''
  JHEP {\bf 0611}, 049 (2006)
  [hep-th/0608088].
   E.~D'Hoker and D.~Z.~Freedman,
  ``Supersymmetric gauge theories and the AdS / CFT correspondence,''
  hep-th/0201253.
  O.~Aharony, A.~B.~Clark and A.~Karch,
  ``The CFT/AdS correspondence, massive gravitons and a connectivity index conjecture,''
  Phys.\ Rev.\ D {\bf 74}, 086006 (2006)
  [hep-th/0608089].

 \bibitem{GMDM-ddnew}
  Y.~J.~Kang and H.~M.~Lee,
``Lightening Gravity-Mediated Dark Matter,''
Eur. Phys. J. C \textbf{80} (2020) no.7, 602
doi:10.1140/epjc/s10052-020-8153-x
[arXiv:2001.04868 [hep-ph]].


\bibitem{XENON:2023cxc}
E.~Aprile \textit{et al.} [XENON],
``First Dark Matter Search with Nuclear Recoils from the XENONnT Experiment,''
Phys. Rev. Lett. \textbf{131} (2023) no.4, 041003
doi:10.1103/PhysRevLett.131.041003
[arXiv:2303.14729 [hep-ex]].

\bibitem{PandaX:2024qfu}
Z.~Bo \textit{et al.} [PandaX],
``Dark Matter Search Results from 1.54 Tonne$\cdot$Year Exposure of PandaX-4T,''
[arXiv:2408.00664 [hep-ex]].

\bibitem{XENON:2018voc}
E.~Aprile \textit{et al.} [XENON],
Phys. Rev. Lett. \textbf{121} (2018) no.11, 111302
doi:10.1103/PhysRevLett.121.111302
[arXiv:1805.12562 [astro-ph.CO]].
  
\bibitem{Aprile:2017iyp}
  E.~Aprile {\it et al.} [XENON Collaboration],
  ``First Dark Matter Search Results from the XENON1T Experiment,''
  Phys.\ Rev.\ Lett.\  {\bf 119} (2017) no.18,  181301
  doi:10.1103/PhysRevLett.119.181301
  [arXiv:1705.06655 [astro-ph.CO]].


\bibitem{Akerib:2016vxi}
  D.~S.~Akerib {\it et al.} [LUX Collaboration],
  ``Results from a search for dark matter in the complete LUX exposure,''
  Phys.\ Rev.\ Lett.\  {\bf 118} (2017) no.2,  021303
  doi:10.1103/PhysRevLett.118.021303
  [arXiv:1608.07648 [astro-ph.CO]].



\bibitem{Cui:2017nnn}
  X.~Cui {\it et al.} [PandaX-II Collaboration],
  ``Dark Matter Results From 54-Ton-Day Exposure of PandaX-II Experiment,''
  Phys.\ Rev.\ Lett.\  {\bf 119} (2017) no.18,  181302
  doi:10.1103/PhysRevLett.119.181302
  [arXiv:1708.06917 [astro-ph.CO]].


\bibitem{Dillon:2016fgw}
B.~M.~Dillon and V.~Sanz,
Phys. Rev. D \textbf{96} (2017) no.3, 035008
doi:10.1103/PhysRevD.96.035008
[arXiv:1603.09550 [hep-ph]].
\bibitem{dileptonsAdam} 
P.~Banerjee, A.~Martin and V.~Sanz,
``Distinguishing among Technicolor/Warped Scenarios in Dileptons,''
JHEP \textbf{01} (2012), 092
doi:10.1007/JHEP01(2012)092
[arXiv:1110.2220 [hep-ph]].

\bibitem{ATLAS:2019erb}
G.~Aad \textit{et al.} [ATLAS],
``Search for high-mass dilepton resonances using 139 fb$^{-1}$ of $pp$ collision data collected at $\sqrt{s}=$13 TeV with the ATLAS detector,''
Phys. Lett. B \textbf{796} (2019), 68-87
doi:10.1016/j.physletb.2019.07.016
[arXiv:1903.06248 [hep-ex]].
\bibitem{ATLAS:2020fry}
G.~Aad \textit{et al.} [ATLAS],
``Search for heavy diboson resonances in semileptonic final states in pp collisions at $\sqrt{s}=13$ TeV with the ATLAS detector,''
Eur. Phys. J. C \textbf{80} (2020) no.12, 1165
doi:10.1140/epjc/s10052-020-08554-y
[arXiv:2004.14636 [hep-ex]].

\bibitem{ATLAS:2022hwc}
G.~Aad \textit{et al.} [ATLAS],
``Search for resonant pair production of Higgs bosons in the $b\bar{b}b\bar{b}$ final state using $pp$ collisions at $\sqrt{s}$ = 13 TeV with the ATLAS detector,''
Phys. Rev. D \textbf{105} (2022) no.9, 092002
doi:10.1103/PhysRevD.105.092002
[arXiv:2202.07288 [hep-ex]].
\bibitem{CMS:2019gwf}
A.~M.~Sirunyan \textit{et al.} [CMS],
``Search for high mass dijet resonances with a new background prediction method in proton-proton collisions at $\sqrt{s} =$ 13 TeV,''
JHEP \textbf{05} (2020), 033
doi:10.1007/JHEP05(2020)033
[arXiv:1911.03947 [hep-ex]].
\bibitem{LZ}
J.~Aalbers \textit{et al.} [LZ Collaboration],
``Dark Matter Search Results from 4.2 Tonne-Years of Exposure of the LUX-ZEPLIN (LZ) Experiment,''
[arXiv:2410.17036 [hep-ex]].


\bibitem{GMDM-dd}
  A.~Carrillo-Monteverde, Y.~J.~Kang, H.~M.~Lee, M.~Park and V.~Sanz,
  ``Dark Matter Direct Detection from new interactions in models with spin-two mediators,''
  JHEP {\bf 1806} (2018) 037
  doi:10.1007/JHEP06(2018)037
  [arXiv:1803.02144 [hep-ph]].
  
  


\bibitem{LZCollaboration:2024lux}
J.~Aalbers \textit{et al.} [LZ Collaboration],
``Dark Matter Search Results from 4.2 Tonne-Years of Exposure of the LUX-ZEPLIN (LZ) Experiment,''
[arXiv:2410.17036 [hep-ex]].



\end{thebibliography}
\end{document}